\documentclass[twocolumn,showpacs,preprintnumbers,amsmath,amssymb,nofootinbib,superscriptaddress,reprint]{revtex4}
\usepackage{amsfonts}
\usepackage{amsmath}
\usepackage{amssymb}
\usepackage{graphicx}
\usepackage{bm}
\usepackage[usenames,dvipsnames]{color}

\begin{document}

\title{Controlling complex networks: How much energy is needed?}

\author{Gang Yan}
\affiliation{Temasek Laboratories, National University of
Singapore, 117411, Singapore}

\author{Jie Ren}
\affiliation{Department of Physics and Centre for Computational Science and Engineering, National University of
Singapore, 117542, Singapore}

\author{Ying-Cheng Lai}
\affiliation{School of Electrical, Computer and Energy Engineering,
Department of Physics, Arizona State University, Tempe, AZ 85287, USA}

\author{Choy-Heng Lai}
\affiliation{Department of Physics and Centre for Computational Science and Engineering, National University of
Singapore, 117542, Singapore}

\author{Baowen Li}
\affiliation{Department of Physics and Centre for Computational Science and Engineering, National University of
Singapore, 117542, Singapore}
\affiliation{Center for Phononics and Thermal Energy Science, Department of Physics,
Tongji University, 200092, Shanghai, China}

\date{\today}

\begin{abstract}

The outstanding problem of controlling complex networks is relevant to many
areas of science and engineering, and has the potential to generate technological
breakthroughs as well. We address the physically important issue of the
energy required for achieving control by deriving and validating scaling
laws for the lower and upper energy bounds. These bounds represent a
reasonable estimate of the energy cost associated with control, and provide
a step forward from the current research on controllability toward ultimate
control of complex networked dynamical systems.

\end{abstract}

\pacs{89.75.-k, 89.75.Fb}
\maketitle

Complex networks are ubiquitous in natural, social, and man-made systems, such as gene regulatory networks,
social networks, mobile sensor networks and so on \cite{Barabasi:review}. A network is composed of nodes and edges. The nodes represent
individual units (e.g., genes, persons, sensors) and the edges represent connections or interactions between the nodes.
The state of a node (e.g., protein being expressed, opinion of a person, position of a sensor) normally evolves over time. And the
evolution depends not only on the node's intrinsic dynamics but also on the couplings with its nearest neighbors \cite{Newman:book}.

On one hand, the couplings between nodes increase the complexity of collective behaviors, which stimulates much interest of modeling, analyzing,
and predicting dynamical processes on complex networks \cite{DynamicalBook}. On the other hand, one may utilize the couplings to control a whole network,
i.e., steering a network from any initial state (vector) to a desired final state, by driving only a few suitable nodes with external signals. In this direction there are good attempts recently from physics \cite{PRL1997,controlRMP,LH:2007,PhysicaD2010,LSB:2011,CowanarXiv,WangarXiv}, biology \cite{biologycontrolbook1,RajaBiology} and engineering \cite{networkcontrolbook2,Network_Control,LCWX:2008,RJME:2009} research communities. Among others, Liu \emph{et al.} studied the controllability of various real-world networks, i.e., the ability to steer a complex network as measured by the minimum number of driver nodes. A main result was that the number of driver nodes required for full control is determined by the network's degree distribution \cite{LSB:2011}. Issues such as achieving control by using only one controller \cite{RJME:2009,CowanarXiv} and making structural
perturbations to the network to minimize the number of control inputs \cite{WangarXiv} have also been addressed.

When control a complex network, an important and unavoidable issue is the cost of control. For instance, in order to control a social network some efforts has
to be devoted to change a few individuals' opinions, while to control an electronic or a mechanical network, some energy has to be consumed to drive a
few elements. Even if a network is controllable in principle, it may not be controllable in practice if it costs an infinite amount of energy or if it requires
too much time to achieve the control.
In this Letter, we address this outstanding issue of \emph{energy cost}, i.e., the amount of efforts or energy that are necessary to produce external signals for steering a complex network, and focus on its {\em lower} and {\em upper} bounds. Suppose a complex network is deemed to be controlled to a desired state in finite time $T_f$, our main results [see Eqs.~(\ref{result}) and (\ref{upperresult})] show the scaling laws of the energy cost bounds with the control time $T_f$ in two different regimes separated by the characteristic time. The results give faithful estimates for the required energy and thus can provide significant insights into bridging network controllability with actual control.

To be able to analyze the energy cost, we study linear networked systems
subject to control inputs. This is the currently standard framework, upon which
the network controllability analysis is built \cite{LH:2007,RJME:2009,LSB:2011,CowanarXiv,WangarXiv}.
A typical system of $N$ nodes and $M$ controllers can be written as
\begin{equation} \label{system2}
\dot{\mathbf{x}}_t = \mathbf{A}\mathbf{x}_{t}
+ \mathbf{B}\mathbf{u}_{t},
\end{equation}
where $\mathbf{x}_{t} = [x_1(t),x_2(t),\ldots, x_N(t)]^\text{T}$
is the state vector of nodes, $\mathbf{u}_{t} = [u_1(t), u_2(t),\ldots,u_M(t)]^\text{T}$
is the input vector of external signals, $\mathbf{B} = \{b_{im}\}$ is the $N\times M$ input
matrix with $b_{im} = 1$ if controller $m$ connects to node $i$ and $b_{im} = 0$ otherwise,
$\mathbf{A} = \{a_{ij}\}$ is the weighted network's adjacency matrix including linear nodal dynamics
$\{a_{ii}\}$.

The typical situation of controlling a complex dynamical network can
be characterized as using external signals $\mathbf{u}_t$ to direct the system
Eq.~(\ref{system2}) from an arbitrary initial state $\mathbf{x}_0$
toward an arbitrary desired state $\mathbf{x}_{T_f}$ in the time
interval $t\in[0,T_f]$. Assuming that the networked system is controllable \cite{LSB:2011,supplement},
our goal is to obtain analytic estimate of the energy cost required
for achieving control, which is defined as \cite{controlbook1}
$\mathcal{E}(T_f)\equiv\int_0^{T_{f}} \|\mathbf{u}_{t}\|^2dt$.
Generally, an infinite number of possibilities exist for choosing
the control input $\mathbf{u}_{t}$ to steer the system Eq.~(\ref{system2})
from $\mathbf{x}_0$ to $\mathbf{x}_{T_f}$. Of all the possible inputs,
the optimal control input is given by
$\mathbf{u}_{t} = \mathbf{B}^\text{T}e^{\mathbf{A}^\text{T}(T_f-t)}\mathbf{W}_{T_f}^{-1}\mathbf{v}_{T_f}$,
which minimizes the energy cost \cite{controlbook1,systembiology}.
The corresponding minimized energy cost is then
$\mathcal{E}(T_f) = \mathbf{v}^\text{T}_{T_f}\mathbf{W}_{T_f}^{-1}\mathbf{v}_{T_f}$,
where
$\mathbf{W}_{T_f} \equiv \int_0^{T_f}e^{\mathbf{A}t}\mathbf{B}\mathbf{B}^\text{T}e^{\mathbf{A}^\text{T}t}dt$
and
$\mathbf{v}_{T_f}\equiv\mathbf{x}_{T_f} - e^{\mathbf{A}T_f}\mathbf{x}_0$
denotes the difference vector between the desired state under control
and the final state during free evolution. For convenience, we set
the origin as the desired state $\mathbf{x}_{T_f} = \mathbf{0}$ and rewrite
the energy cost as
\begin{equation} \label{energy}
\mathcal{E}(T_f) = \mathbf{x}^\text{T}_0\mathbf{H}^{-1}\mathbf{x}_0,
\end{equation}
where
$\mathbf{H}(T_f) \equiv e^{-\mathbf{A}T_f}\mathbf{W}_{T_f}e^{-\mathbf{A}^\text{T}T_f}$
is the symmetric Gramian matrix \cite{controlbook1}. When the system is
controllable, $\mathbf{H}$ is positive-definite (PD), otherwise it is
non-invertible. In the following we focus on the normalized energy cost
\begin{equation}
E(T_f)={\mathcal{E}(T_f)}/{\| \mathbf{x}_0\|^2} =
\frac{\mathbf{x}^\text{T}_0\mathbf{H}^{-1}\mathbf{x}_0}{\mathbf{x}^{\text T}_0\mathbf{x}_0}.
\end{equation}
When $\mathbf{x}_0$ is parallel to the direction of one of
$\mathbf{H}$'s eigenvectors, the corresponding inverse of the eigenvalue
has the physical meaning of normalized energy cost associated with
controlling the system along the particular eigendirection. Using the
Rayleigh-Ritz theorem \cite{matrixbook}, we can bound the normalized
energy cost as
\begin{equation} \label{bound}
\frac{1}{\eta_{\text{max}}}\equiv E_{\text{min}}\leq E(T_f)
\leq E_{\text{max}}\equiv \frac{1}{\eta_{\text{min}}},
\end{equation}
where $\eta_{\text{max}}$ and $\eta_{\text{min}}$ are the maximal and minimal
eigenvalues of the PD matrix $\mathbf{H}$, respectively.

To proceed, we focus on the lower and upper bounds of normalized energy
cost for the case of single-node control. To analytically calculate the
quantities $1/\eta_{\text{max}}$ and $1/\eta_{\text{min}}$, for weighted undirected networks,
we decompose the matrix $\mathbf{A}$ in terms of its eigenvectors as
$\mathbf{A=VSV}^{\text{T}}$, where $\mathbf{V}$ is the orthonormal
eigenvector matrix that satisfies $\mathbf{VV}^{\text{T}}=\mathbf{V}^{\text{T}}\mathbf{V}=\mathbf{I}$,
$\mathbf{S} = \mathrm{diag}\{\lambda_1, \lambda_2, \ldots, \lambda_N\}$ with
descending order
$\lambda_1>\lambda_2>\ldots>\lambda_N$. We thus have
$e^{\mathbf{A}t} = e^{\mathbf{A}^\text{T}t} = \mathbf{V}e^{\mathbf{S}t}\mathbf{V}^{\text{T}}$. Substituting these expressions into
the Gramian matrix and noting that $\mathbf{V}$ is time-independent, we have
\begin{equation} \label{H}
\mathbf{H}= \mathbf{V}e^{-\mathbf{S}T_f}(\int_0^{T_f} e^{\mathbf{S}t}\mathbf{V}^{\text{T}}\mathbf{B}\mathbf{B}^\text{T}\mathbf{V}e^{\mathbf{S}t}dt)e^{-\mathbf{S}T_f}\mathbf{V}^{\text{T}}.
\end{equation}
Denoting the only node under direct control as $c$, we have that $\mathbf{B}$ is an
$N\times 1$ matrix, of which all elements are zeros except the $c$th element, which
is one. After some amount of algebra, we obtain
\begin{equation} \label{finalH}
H_{ij}=\sum_{\alpha=1}^N\sum_{\beta=1}^N\frac{V_{i\alpha}V_{c\alpha}V_{c\beta}V_{j\beta}}
{\lambda_{\alpha} + \lambda_{\beta}}
\left(1 - e^{-\left(\lambda_{\alpha} + \lambda_{\beta}\right)T_f}\right),
\end{equation}
where the Roman letters $i,j,c$ are node indices in the real space while the
Greek letters $\alpha, \beta$ are running indices in the eigenspace.

To carry the analysis further, we note that there are two distinct regimes
in terms of the control time $T_f$. In the {\em small} $T_f$ regime where
$T_f \ll 1/|\lambda_{\alpha} + \lambda_{\beta}|$, we can expand
$e^{-(\lambda_{\alpha} + \lambda_{\beta})T_f}\approx1-(\lambda_{\alpha} + \lambda_{\beta})T_f$
and obtain
$H_{ij} \approx T_f\sum_{\alpha=1}^N\sum_{\beta=1}^N
V_{i\alpha}V_{c\alpha}V_{c\beta}V_{j\beta}=T_c\delta_{ic}\delta_{cj}$.
In this case, we have $H_{ij} \approx 0$ for all $i$ and $j$ except $H_{cc} \approx T_f$
so that the maximal eigenvalue of matrix $\mathbf{H}$ can be approximated
as $T_f$. Consequently, for the small $T_c$ regime, we have
$E_{\text{min}}\equiv1/\eta_{\text{max}} \approx 1/T_f$,
regardless of the form of the matrix $\mathbf{A}$ and of the value of $c$.
In contrast, in the {\em large} $T_f$ regime characterized by
$T_f \gg 1/|\lambda_{\alpha} + \lambda_{\beta}|$, we can
approximate the maximal eigenvalue of $\mathbf{H}$ by its trace, which
has been numerically verified:
$\eta_{\text{max}}\approx\sum^N_{\alpha=1}\eta_{\alpha} \equiv\mathrm{Tr}[\mathbf{H}]
=\sum_{i}^{N}\sum^N_{\alpha}\sum^N_{\beta}
\frac{V_{i\alpha}V_{c\alpha}V_{c\beta}V_{i\beta}}{\lambda_{\alpha} + \lambda_{\beta}}
\left(1 - e^{-(\lambda_{\alpha} + \lambda_{\beta})T_f}\right)
=\sum_{\alpha=1}^N\frac{V_{c\alpha}^2}{2\lambda_{\alpha}}
\left(1 - e^{-2\lambda_{\alpha}T_f}\right)$.
If $\mathbf{A}$ is PD, the term $e^{-2\lambda_{\alpha}T_f}$
vanishes for large $T_f$. We thus have
$E_{\text{min}}\equiv1/\eta_{\text{max}}
\approx 1/\sum_{\alpha=1}^N\frac{V_{c\alpha}^2}{2\lambda_{\alpha}}
\left(1 - e^{-2\lambda_{\alpha}T_f}\right)
\approx1/\sum_{\alpha=1}^N\frac{V_{c\alpha}^2}{2\lambda_{\alpha}}
= 1/[(\mathbf{A}+\mathbf{A}^\text{T})^{-1}]_{cc}$.
Note that, since the matrix $\mathbf{A}$ is independent of $T_f$, the factor
$1/[(\mathbf{A}+\mathbf{A}^\text{T})^{-1}]_{cc}$ is time-independent too. This means
that, when $\mathbf{A}$ is PD, the lower bound of the energy cost converges to a
constant value for large $T_f$. If $\mathbf{A}$ is not PD, i.e., at least one
of $\mathbf{A}$'s eigenvalues is negative, the most negative eigenvalue $\lambda_N$
will dominate the behavior of $\mathbf{H}$:
$H_{ij}\approx\frac{V_{iN}V_{cN}^2V_{jN}}{2\lambda_N}\left(1 - e^{-2\lambda_NT_f}\right)\sim e^{-2\lambda_NT_f}$.
As a result, the maximal eigenvalue of $\mathbf{H}$ grows exponentially with $T_f$:
$\eta_{\text{max}} \sim e^{-2\lambda_NT_f}$ so that
$E_{\text{min}} \sim e^{2\lambda_NT_f}$.
Since $\lambda_N < 0$, the lower bound of the energy cost vanishes exponentially
with the control time $T_f$. In the borderline case where $\mathbf{A}$ is semi PD, i.e.,
$\lambda_{\alpha} > 0$ for $\alpha = 1, 2, \ldots, N-1$ and $\lambda_{N} = 0$,
the behavior of $\mathbf{H}$ can be characterized as:
$H_{ij} \approx \lim_{\lambda_{N} \rightarrow 0}
\frac{V_{iN}V_{cN}^2V_{jN}}{2\lambda_N}\left(1 - e^{-2\lambda_NT_f}\right)\sim T_f^{-1}$.

\begin{figure}
\begin{center}
\includegraphics[width=0.49\columnwidth]{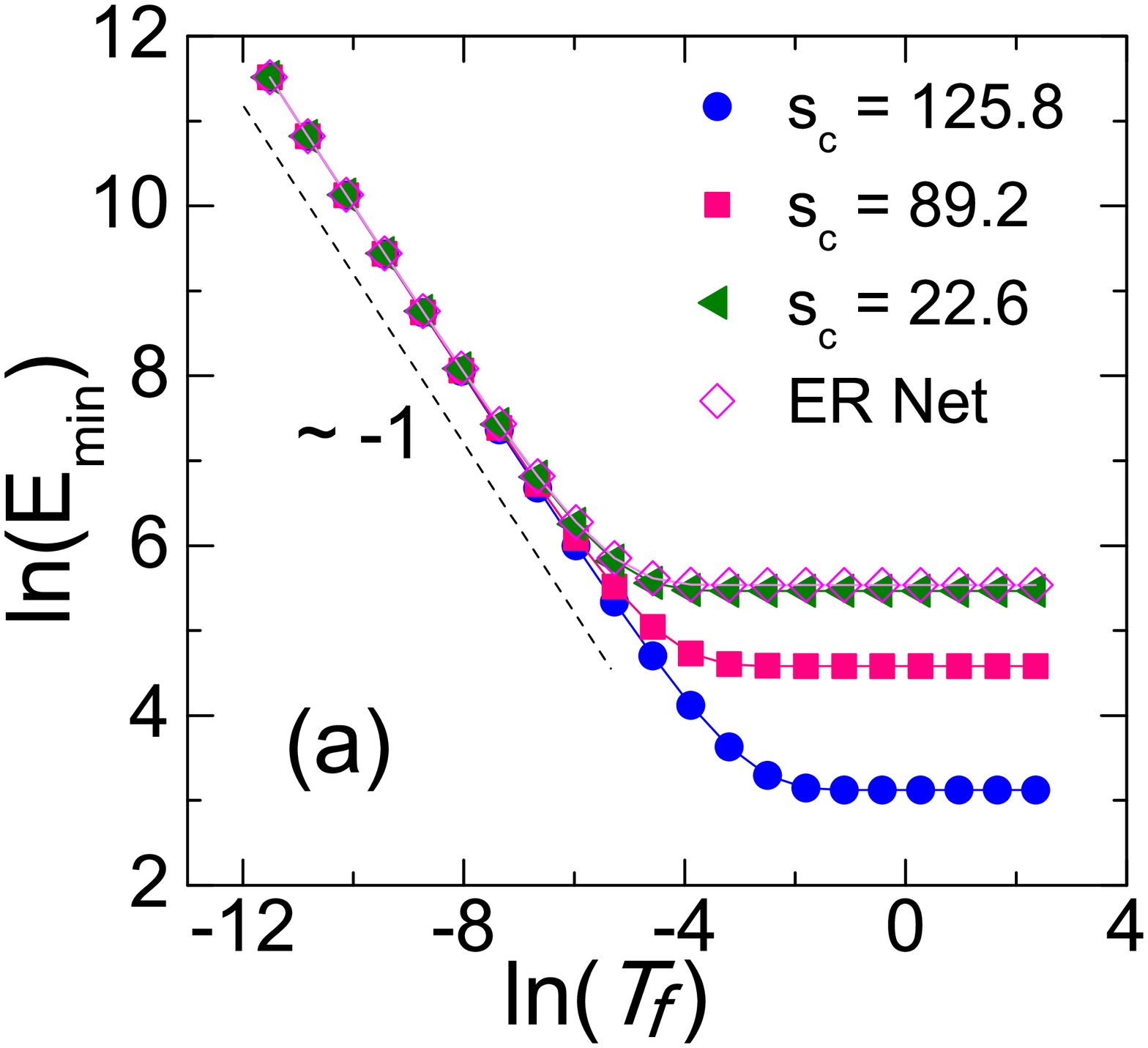}
\includegraphics[width=0.49\columnwidth]{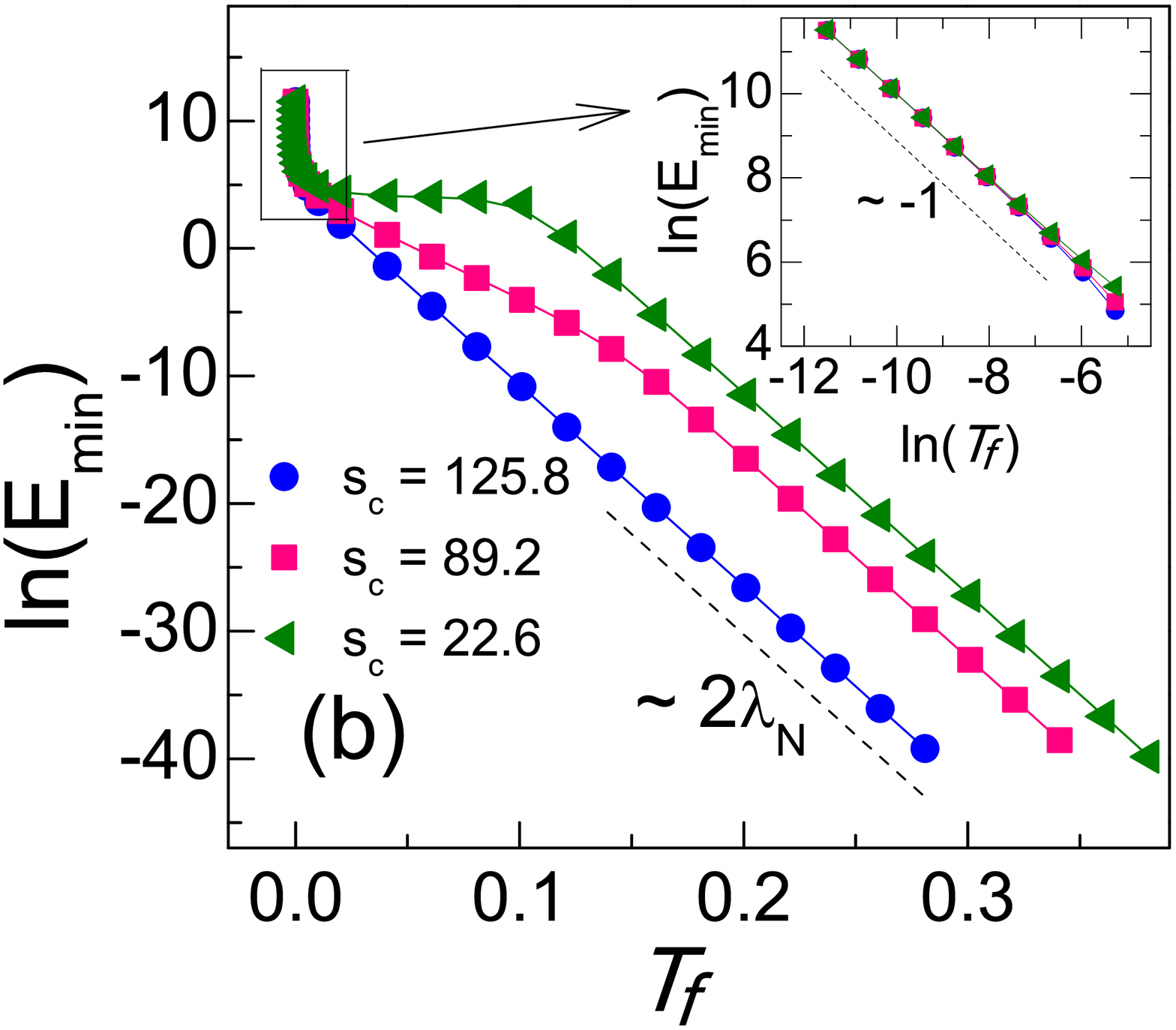}
\end{center}\vspace{-5mm}
\caption{(color online). Lower bound of the energy cost
$E_{\text{min}} \equiv 1/\eta_{\text{max}}$ versus the control time $T_f$.
All networks are weighted BA scale-free networks except one weighted ER random network in (a), with the same size $N = 500$
and $\langle s\rangle = 20$. The $s_c$ denotes the strength of directly controlled node. In (a) $a = -150$
which makes $\mathbf{A}$ PD. In (b) $a = -50$ thus
$\mathbf{A}$ is not PD. The dashed line in the semi-log plot
in (b) has a slope $2\lambda_N$. The symbols represent
the same quantities calculated numerically and the solid lines represent
the results from the estimation $\eta_{\text{max}} \approx \text{Tr}[\mathbf{H}]$.}
\label{figlower}
\end{figure}

Our theoretical estimates for the \emph{lower bound} $E_{\text{min}}$ of the energy cost can be summarized as
\begin{equation} \label{result}
E_{\text{min}}
  \begin{cases}
     \approx T_f^{-1} & \text{small $T_f$} \\
     \approx \frac{1}{[(\mathbf{A}+\mathbf{A}^\text{T})^{-1}]_{cc}} & \text{large $T_f$, $\mathbf{A}$ is PD} \\
     \xrightarrow[\sim \exp{\left(2\lambda_NT_f\right)}]{\sim~T_f^{-1}}0  & \text{large $T_f$, $\mathbf{A}$ is } \frac{\text{semi PD}}{\text{not PD}}
  \end{cases}.
\end{equation}
Numerical support for Eq.~(\ref{result}) is
shown in Fig. \ref{figlower}. We use scale-free networks
generated by the Barab\'asi-Albert (BA) model \cite{BAmodel} and
Erd\"os-R\'eyni (ER) type of random networks \cite{ERmodel}.
The link weights are randomly generated from the uniform interval $[0.5,1.5]$.
The linear nodal dynamics are set as ${a_{ii}} = -(a+s_i)$ where $s_i = \sum^N_{j=1,j\neq i}a_{ij}$ is the strength of node $i$,
and $a$ is such a tunable parameter that one can conveniently change $\mathbf{A}$ between positive and negative definite. We note
that other node-dependent settings of $a_{ii}$ will not affect our results.
Use the method proposed in \cite{LSB:2011} one can find the weighted network is controllable,
except some pathological link-weights sets of measure zero,
by any single driver node. We numerically compute the lower bound
according to Eqs.~(\ref{bound}) and (\ref{H}).
From Figs.~\ref{figlower}(a) and inset of \ref{figlower}(b), we see that,
for the small $T_f$ regime, $E_{\text{min}}$ decays as a power law $T_f^{-1}$,
regardless of $\mathbf{A}$ and $c$, agreeing with our theoretical result.
In the large $T_f$ regime, the behavior of $E_{\text{min}}$ is determined by
the signs of the eigenvalues of $\mathbf{A}$. In particular, if the eigenvalues
are all positive, the dynamics in the absence of control, i.e.,
$\dot{\mathbf{x}}_{t} = \mathbf{A}\mathbf{x}_{t}$, will force the nodal
states to depart away from the zero state. Thus, even given sufficiently large
time, one has to consume some amount of energy to steer the nodes back.
As shown in Fig.~\ref{figlower}(a),
$E_{\text{min}}$ converges to a constant value as $T_f$ is increased, which
agrees with our predicted value $1/[(\mathbf{A}+\mathbf{A}^T)^{-1}]_{cc}$. In
contrast, if $\mathbf{A}$ is not PD, $E_{\text{min}}$ vanishes exponentially,
as shown in Fig.~\ref{figlower}(b). The corresponding exponent is
$2\lambda_{N}$, which is consistent with our theoretical estimate
in Eq.~(\ref{result}) as well.

\begin{figure}
\begin{center}
\includegraphics[width=0.49\columnwidth]{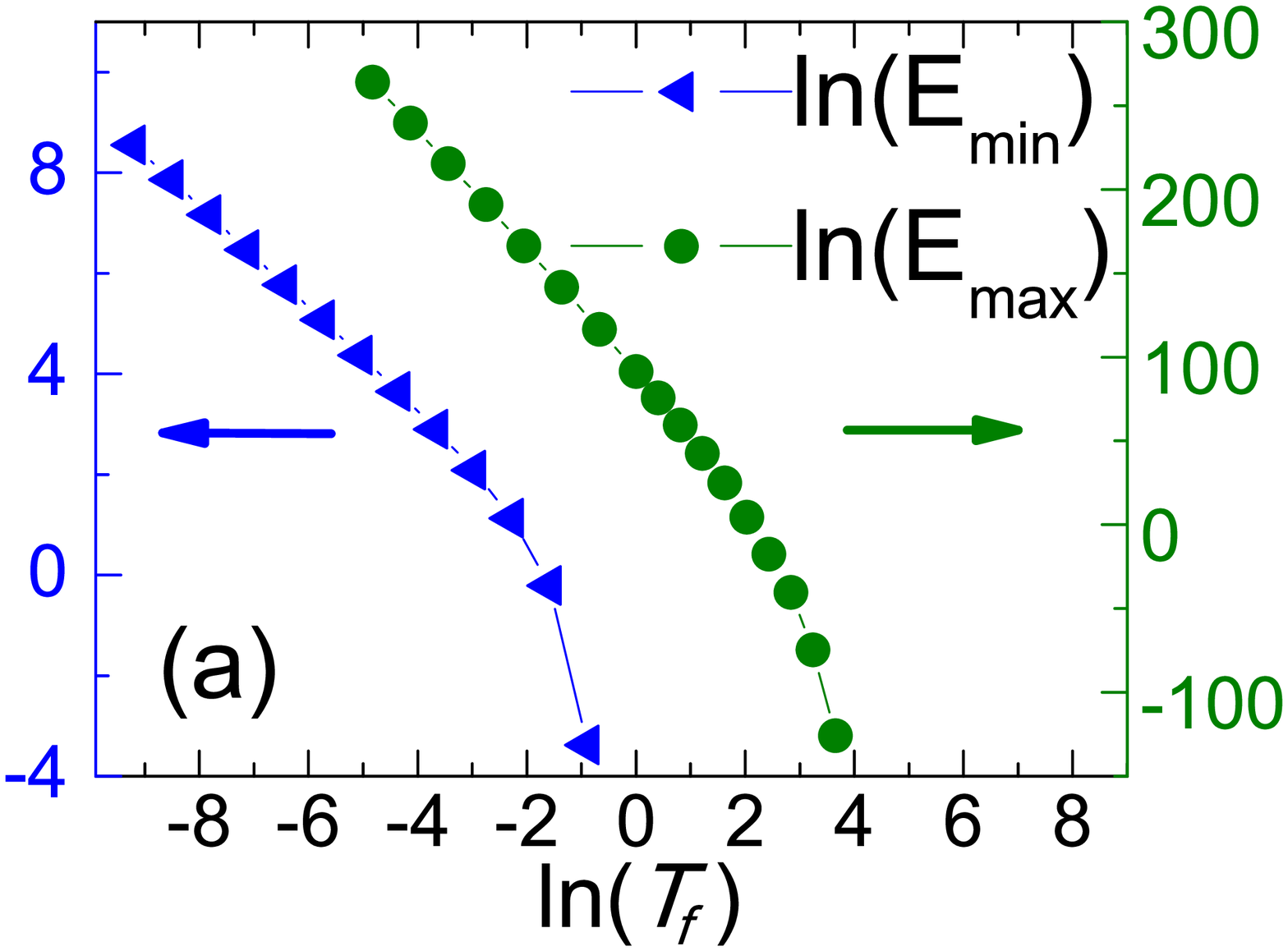}
\includegraphics[width=0.49\columnwidth]{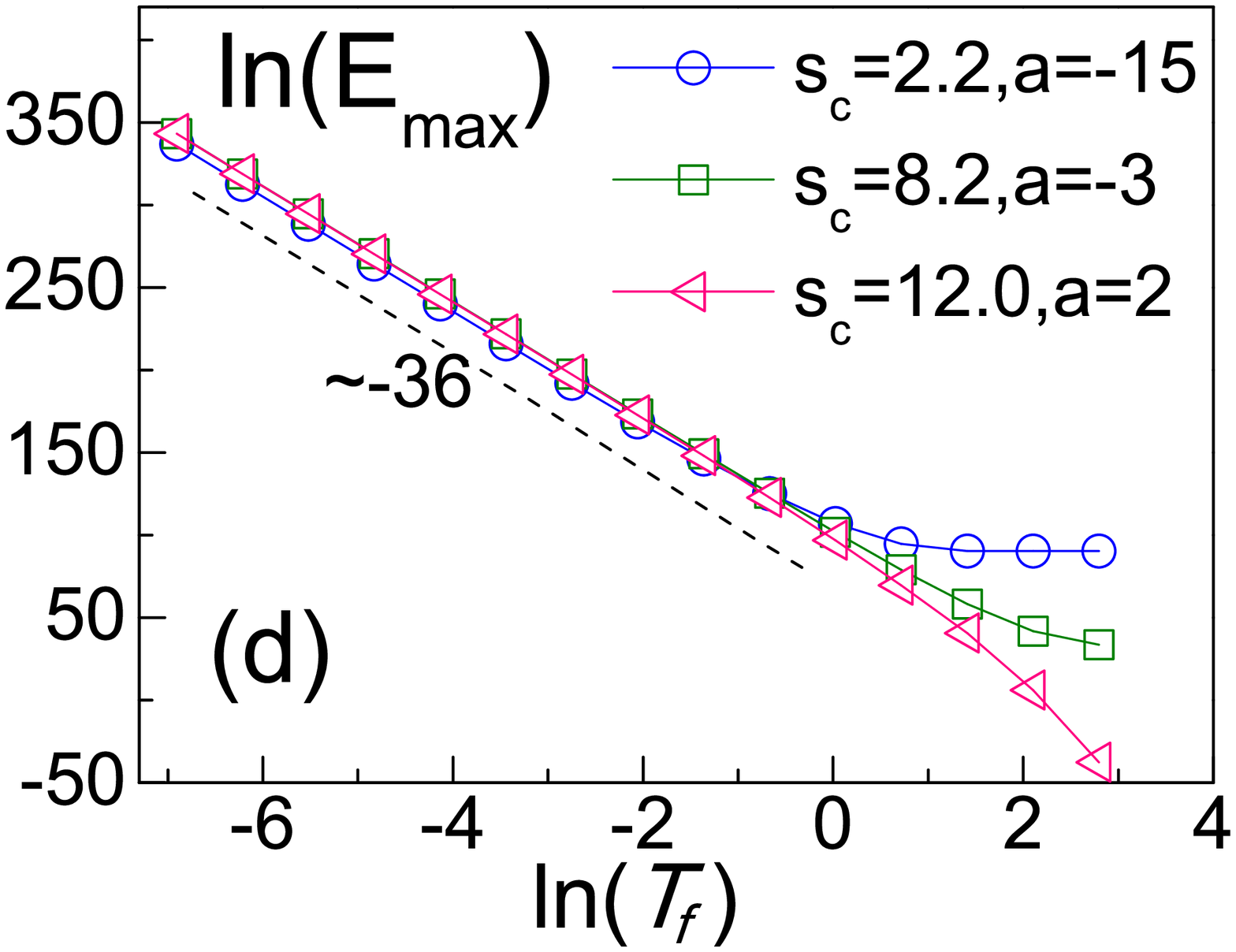}
\includegraphics[width=0.49\columnwidth]{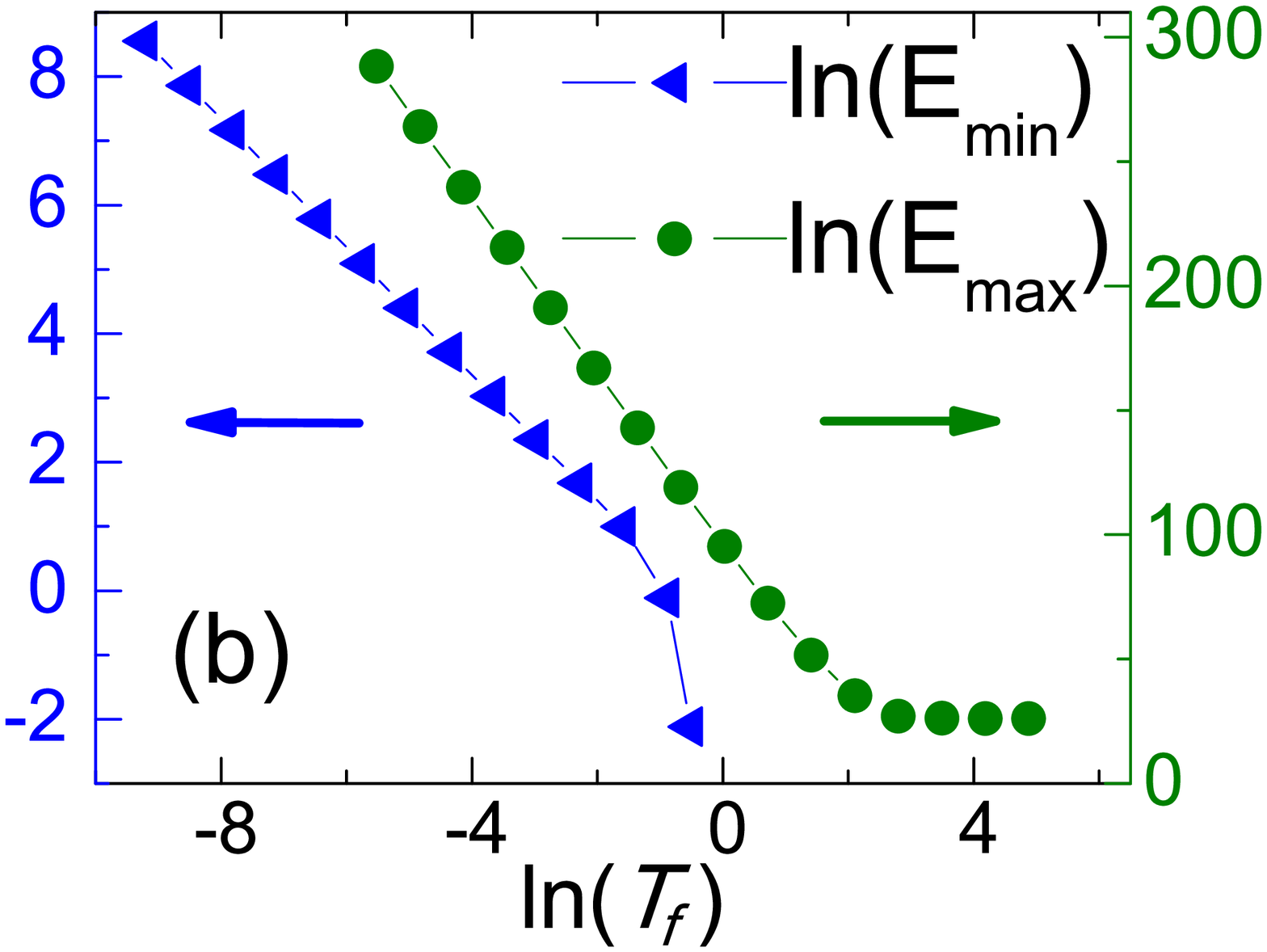}
\includegraphics[width=0.49\columnwidth]{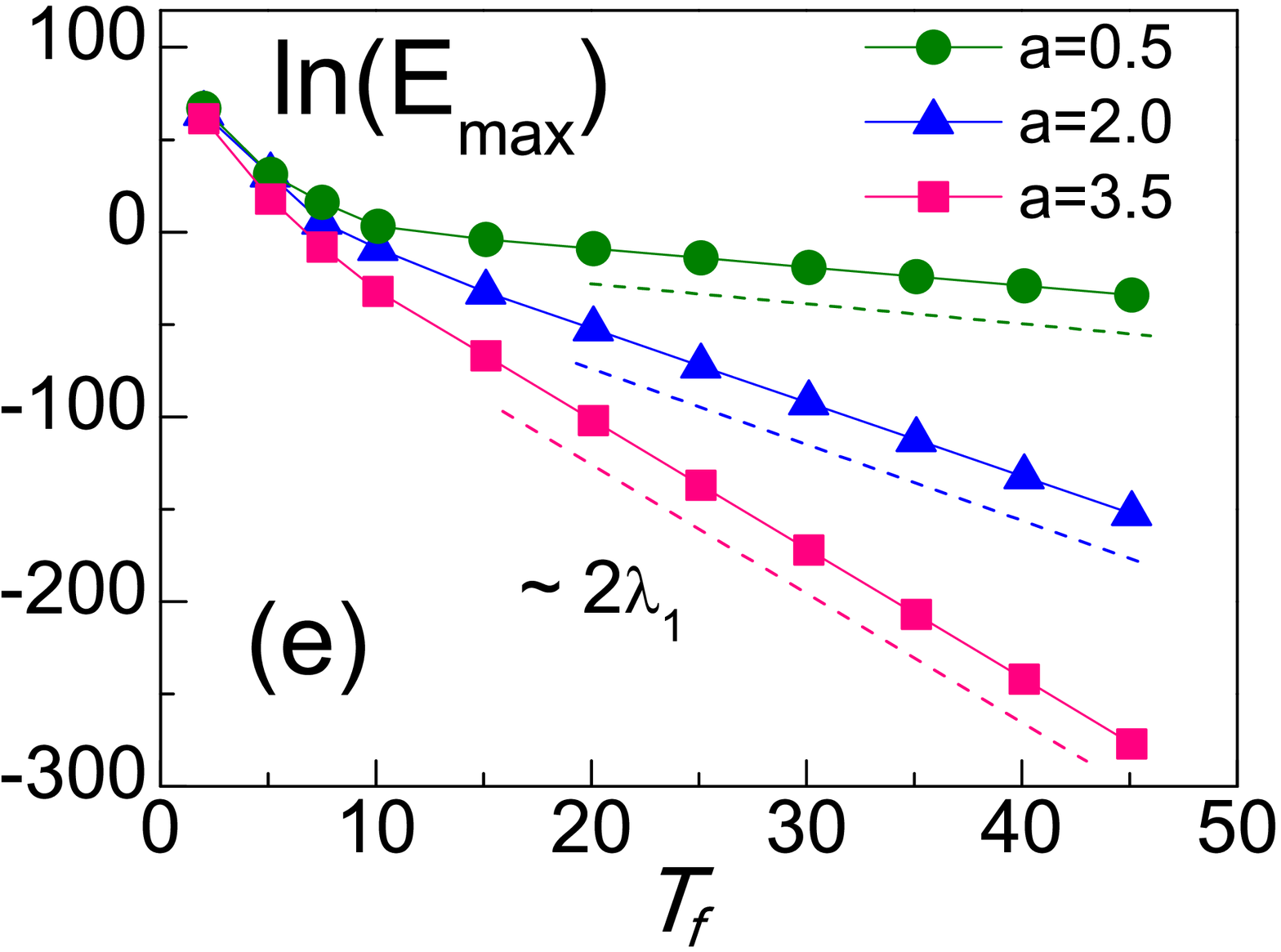}
\includegraphics[width=0.49\columnwidth]{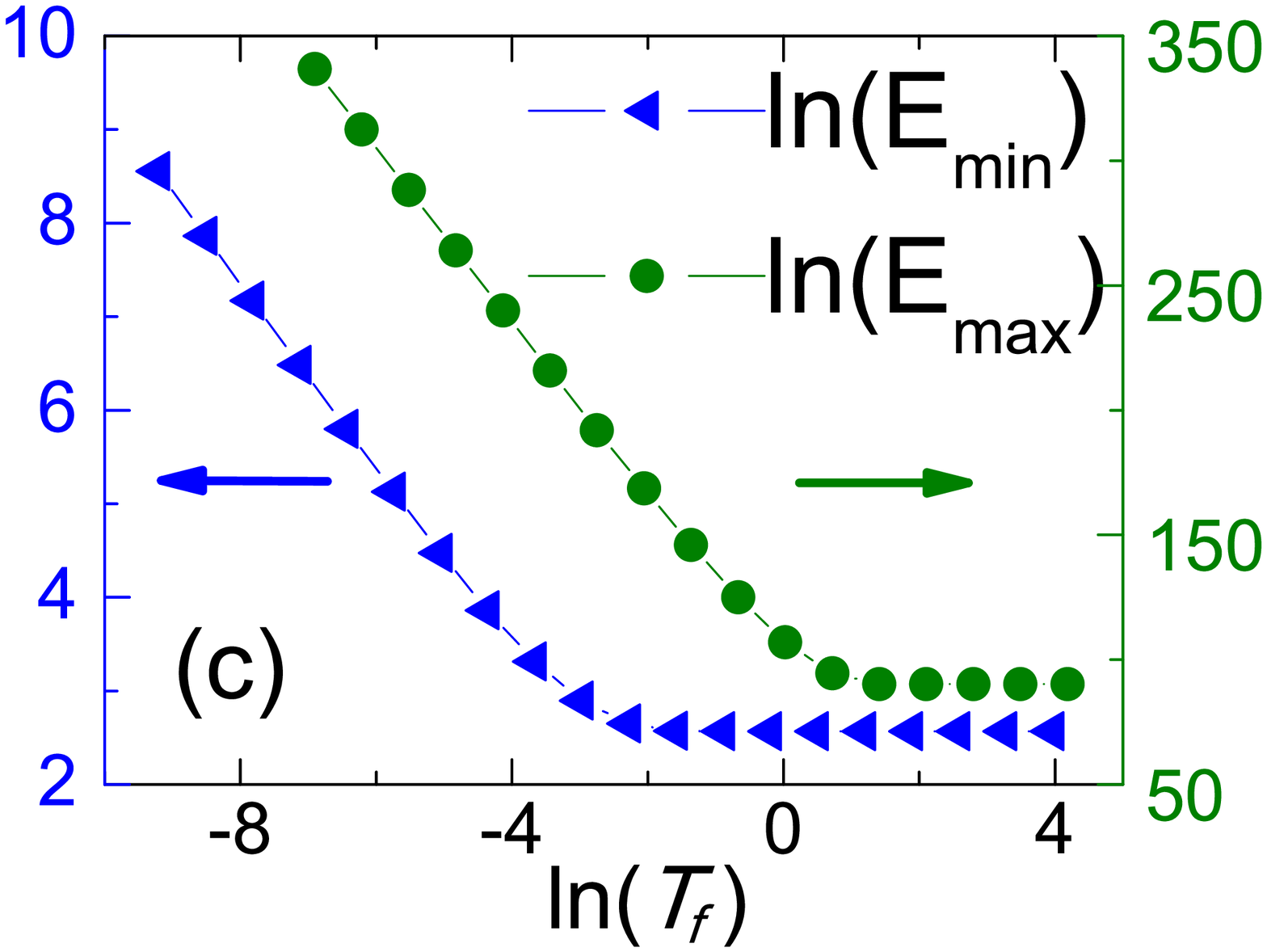}
\includegraphics[width=0.49\columnwidth]{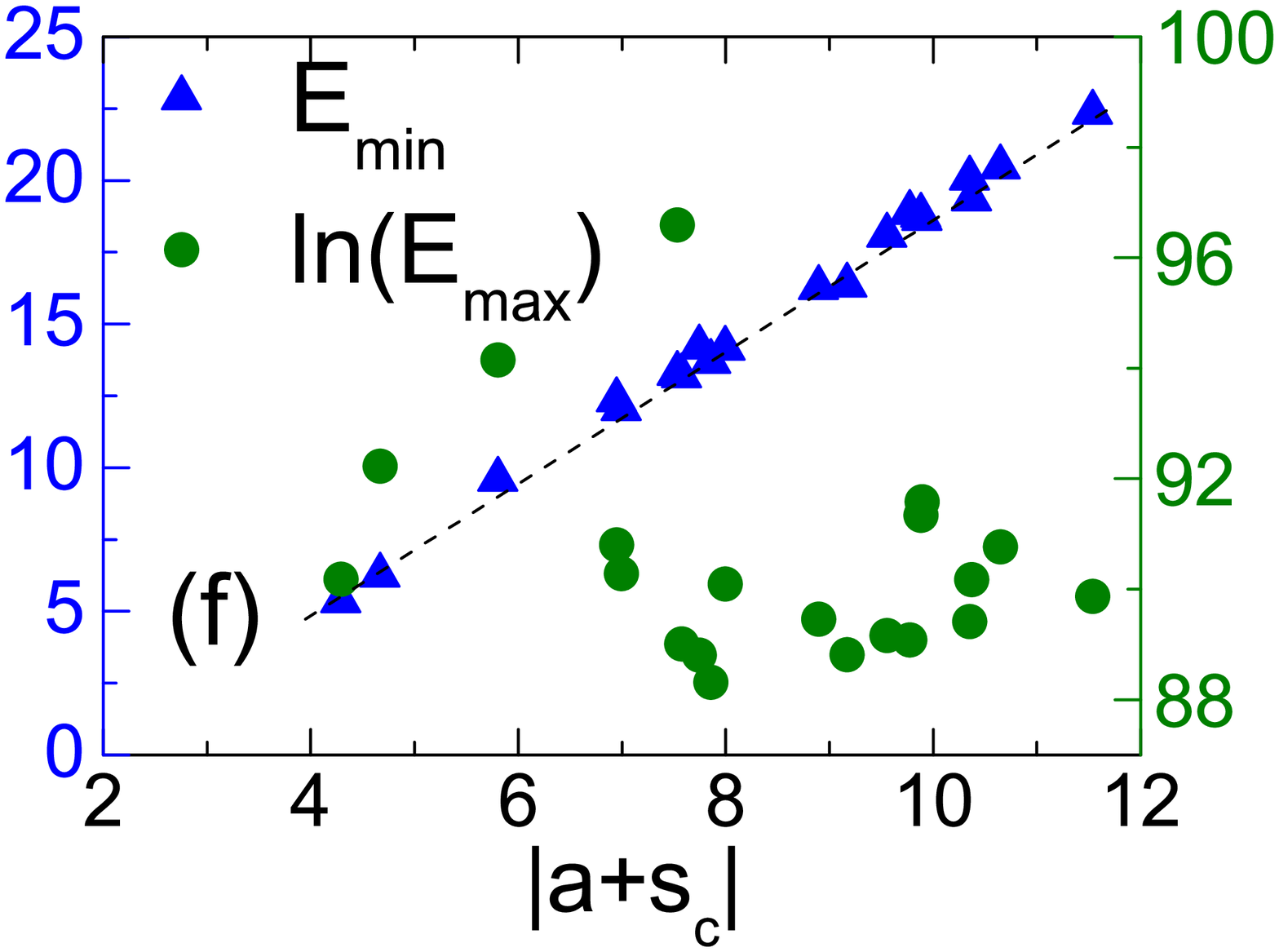}
\end{center}
\caption{(color online). Upper bound of the control energy cost
$E_{\text{max}} \equiv 1/\eta_{\text{min}}$ for a weighted BA network with
20 nodes. In (a), $a = 2$ thus $\mathbf{A}$ is ND. In (b),
$a = -5$. In (c), $a = -20$ so that $\mathbf{A}$ is PD.
In (a-c), ({\color{OliveGreen} \Large$\bullet$}) represent
the upper bound $E_{\text{max}}$ while ({\color{Blue} $\blacktriangleleft$})
represent the corresponding lower bound $E_{\text{min}}$ (included
for comparison). In (d) the decaying behavior of $E_{\text{max}}$ is
shown for different $s_c$ and $a$ values. The dash line has a slope $-36$.
In (e) the exponential decay of $E_{\text{max}}$ for large $T_f$ is plotted for different values of $a$.
The slopes of dashed lines are $2\lambda_{1}$ respectively.
In (f) the constant values of the energy cost in (c) are shown as a
function of $|a+s_c|$. The slope of the dashed line is $2$.}
\label{figupper}
\end{figure}

We now turn to the \emph{upper bound} of the energy cost
$E_{\text{max}}\equiv1/\eta_{\text{min}}$. As indicated by Eq.~(\ref{finalH}),
most elements of the matrix $\mathbf{H}$ are small, especially for the
small $T_f$ regime. Consequently, $\mathbf{H}$ is generally ill-conditioned
\cite{matrixbook} and its minimal eigenvalue is typically very small
(though positive). Thus, to control a large-size network,
$E_{\text{max}}$ can be very large. The underlying
physical reason is that, when only one node is subject to control, the effect
on other nodes will not be direct but instead will be indirect
through various paths on the network. The end result is
that we need to steer the whole system in the state space by following highly
circuitous, though smooth, routes \cite{supplement}, a process that requires
a large amount of energy.

Typical results computed from Eqs.~(\ref{bound}) and (\ref{H}) are shown in
Figs.~\ref{figupper}(a-c). For small
$T_f$, the upper bound $E_{\text{max}}$ exhibits power-law decay, similar to
the behavior of the lower bound, but the decay exponent for $E_{\text{max}}$
assumes a much larger value that is independent of $a$ and $c$ [see Fig.~\ref{figupper}(d)].
For large $T_f$, $E_{\text{max}}$ will converge to a constant value if $\mathbf{A}$
is not negative definite (ND), or will vanish exponentially if $\mathbf{A}$ is ND.
The corresponding exponent is given by $2\lambda_{1}$, where
$\lambda_{1}$ is the least negative eigenvalue of $\mathbf{A}$, as shown in Fig.~\ref{figupper}(e).
This is due to the fact that, in the large $T_f$
limit, the behavior of $H_{ij}^{-1}$ is dominated by the mode with the least negative
eigenvalue $\lambda_{1}$, which contributes the slowest increase to $H_{ij}$.
As a result, we have
$E_{\mathrm{max}}\sim[\mathbf{H}^{-1}]_{ij}\sim H_{ij}^{-1}
\sim\frac{2\lambda_1}{\left(1-\exp{(-2\lambda_1T_f)}\right)}\sim
e^{2\lambda_1T_f}$.
In the borderline case, i.e., $\mathbf{A}$ is semi ND, the upper bound decays according
to $T_f^{-1}$: $E_{\mathrm{max}}\sim \lim_{\lambda_1\rightarrow 0}
\frac{2\lambda_1}{\left(1-\exp{(-2\lambda_1T_f)}\right)}\sim T_f^{-1}$.
Such a behavior in both $E_{\text{max}}$ and $E_{\text{min}}$ has been numerically
verified~\cite{supplement}.

The results for the \emph{upper bound} can be summarized as:
\begin{equation} \label{upperresult}
 E_{\text{max}}
  \begin{cases}
     \approx T_f^{-\theta}\;(\theta \gg 1) & \text{small $T_f$}\\
     = \varepsilon(\mathbf{A},c) & \text{large $T_f$, $\mathbf{A}$ is not ND}\\
     \xrightarrow[\sim \exp{\left(2\lambda_1T_f\right)}]{\sim~T_f^{-1}}0 & \text{large $T_f$, $\mathbf{A}$ is } \frac{\text{semi ND}}{\text{ND}}
  \end{cases},
\end{equation}
where $\varepsilon(\mathbf{A},c)$ denotes a positive value that depends on the matrix
$\mathbf{A}$ and the controlled node $c$. For the constant value of the lower bound as
described in Eq.~(\ref{result}), one may approximate
$1/[(\mathbf{A}+\mathbf{A}^T)^{-1}]_{cc}\approx 2a_{cc}$
so that $E_{\mathrm{min}}$ is proportional to $|a+s_c|$. However, as shown in
Fig.~\ref{figupper}(f), there appears no proportional relationship between the
constant value $\varepsilon(\mathbf{A},c)$ of $E_{\mathrm{max}}$ and $a_{cc}$ of the controlled node. This indicates that directly controlling a node
with larger degree does not generally result in less energy cost.

Actually, when the system matrix $\mathbf{A}$ is PD and the control time $T_f \rightarrow \infty$, Eq.~\ref{finalH} reduces to
$H^{\infty}_{ij} = \sum_{\alpha=1}^N\sum_{\beta=1}^N\frac{V_{i\alpha}V_{c\alpha}V_{c\beta}V_{j\beta}}
{\lambda_{\alpha} + \lambda_{\beta}}$ which is the solution of $\mathbf{AH}^{\infty} + \mathbf{H}^{\infty}\mathbf{A}^{\text{T}} = \mathbf{BB}^{\text{T}}$
and can be naturally interpreted as dynamical correlation~\cite{RWLL:PRL10}, between nodes $i$ and $j$ with respect to controlled (driver) node $c$.
So $\varepsilon(\mathbf{A},c)$ is the inverse of the smallest eigenvalue of the correlation matrix $\mathbf{H}^{\infty}$.
From this point of view, two indications come out immediately: Firstly, to find optimal driver node in a network, one should consider
the node viewing from which the rest nodes are most dissimilar. The reason is that, controlling a central hub node, though may transmit
external signals fast, can induce star-like structure which makes the rest nodes more similar to each other.
When nodes are more structurally similar, they tend to have more similar dynamical correlations with other nodes so that the
corresponding rows in $\mathbf{H}^{\infty}$ become more similar. As a consequence, the smallest eigenvalue of $\mathbf{H}^{\infty}$ will be
less. In other words, we have to consume more energy to independently steer similar nodes in order to
fully control the network. Secondly, for randomized networks, the more heterogeneous the node-degrees, the higher the energy cost
of control, on the average (see Section III of~\cite{supplement}).
 Take randomized BA and ER networks for example, we compare the values of $\varepsilon(\mathbf{A},c)$, i.e., $\varepsilon_{\text{BA}}$ and $\varepsilon_{\text{ER}}$ in Fig.~3(a). It shows that the upper bound of energy cost for controlling BA networks is much larger than that for
 controlling ER networks.

\begin{figure}
\begin{center}
\includegraphics[width=0.49\columnwidth]{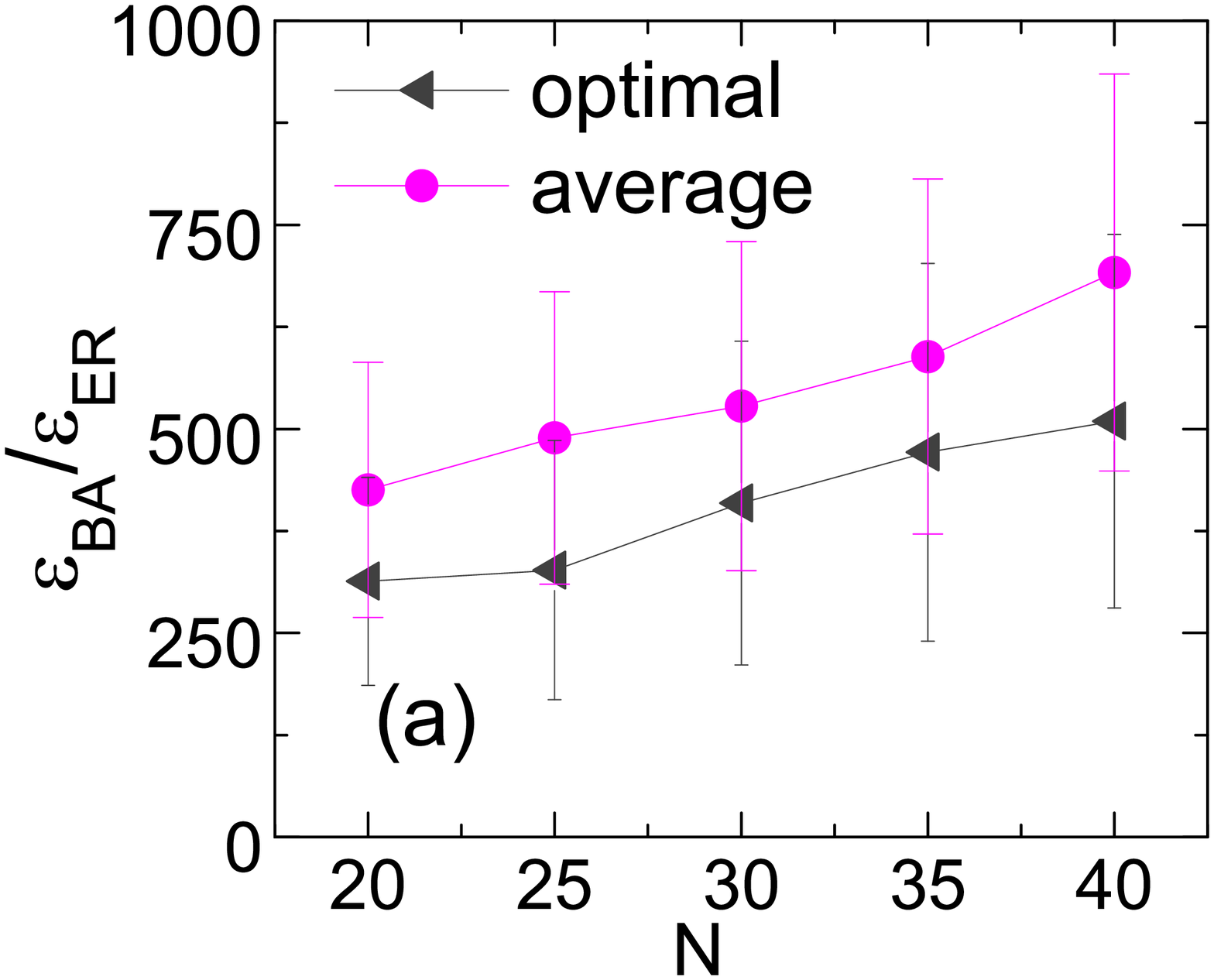}
\includegraphics[width=0.49\columnwidth]{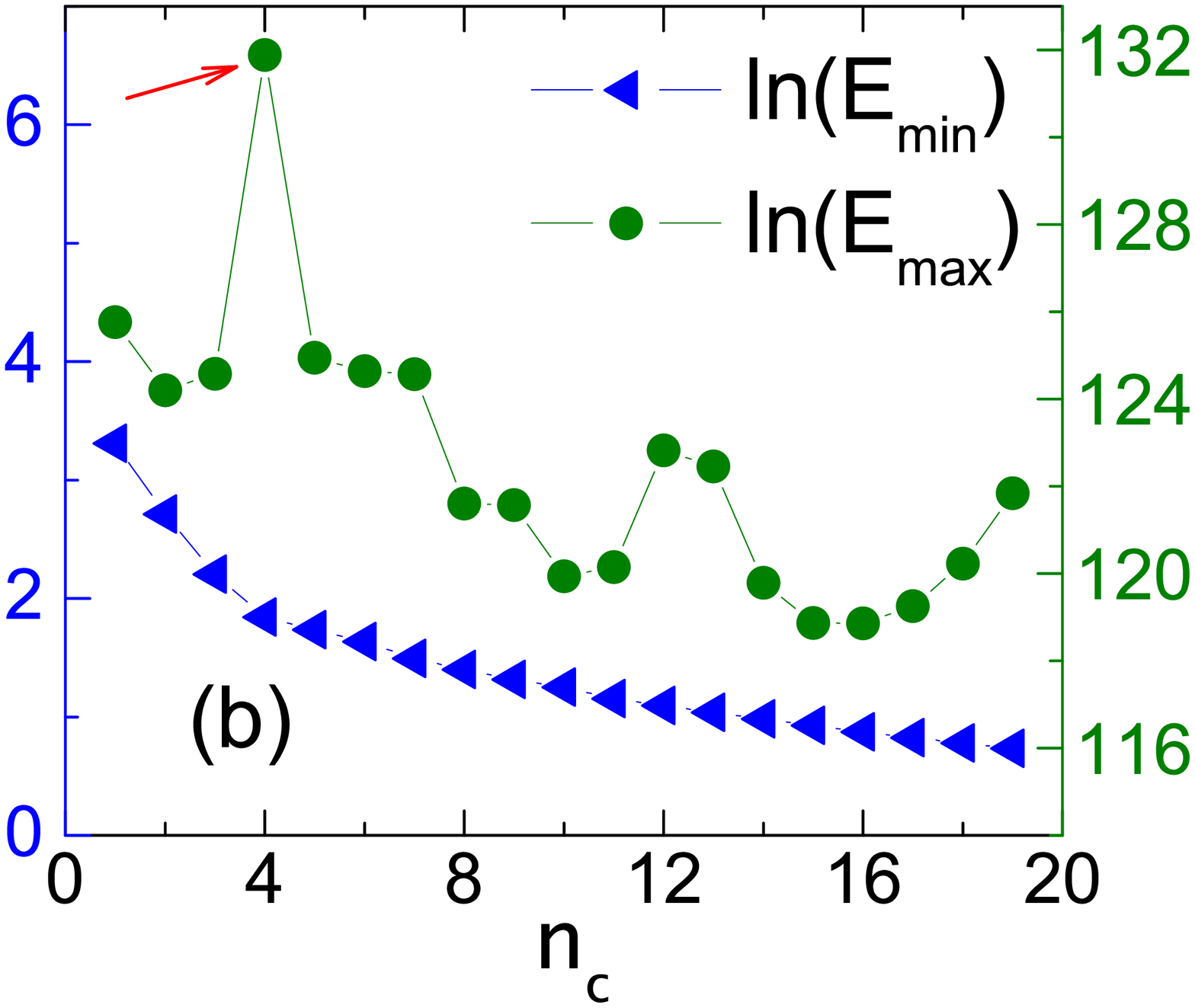}
\end{center}\vspace{-5mm}
\caption{(color online). (a) The ratio $\varepsilon_{BA}/\varepsilon_{ER}$ for different network size $N$.
In order to eliminate the effects of nodal dynamics and strength, we fix the values of $a_{ii}$ and $\langle s\rangle$.
The results include the ratio for optimal driver node ({\color{Gray} $\blacktriangleleft$}) and the ratio of averaging over
different driver nodes ({\color{Magenta} \Large$\bullet$}). The error bars are caused by different generations of network topology and link weights.
(b) $E_{\text{min}}$ ({\color{Blue} $\blacktriangleleft$}, left) and $E_{\text{max}}$ ({\color{OliveGreen} \Large$\bullet$}, right) versus
$n_c$, the number of directly controlled nodes. The dot pointed by the arrow corresponds to the node with largest degree
in the network.}
\label{EvsNc}
\end{figure}
We have also studied the energy cost associated with the control scheme
proposed in a recent work \cite{CowanarXiv}, i.e., controlling more than
one node by a common controller. Fig.~\ref{EvsNc}(b) shows the effect of $n_c$, the
number of directly controlled nodes, on the energy cost, which reveals that controlling
more nodes will induce smaller value of the lower energy bound. This, however,
does not hold for the upper bound. In fact, adding a node with large degree
into the directly controlled node-set may drastically increase
the energy cost. This result is consistent, to a certain degree, with that found
in Ref.~\cite{LSB:2011} which shows the driver nodes tend to avoid the
high-degree nodes.

It is noteworthy that our results can be easily generalized to weighted \emph{directed} networks. If a network is controllable
by one driver node, the eigenvalues of the corresponding system matrix $\mathbf{A}$ are non-degenerate~\cite{RJME:2009} though may be not all real.
Thus we have $\mathbf{A} = \mathbf{VSV}^{-1}$ where $\mathbf{S} = \text{diag}\{\lambda_1,\lambda_2,\ldots,\lambda_N\}$ with descending order of the real part $\text{Re}\lambda_1 \geq \text{Re}\lambda_2
\geq \ldots \geq\text{Re}\lambda_N$. Similarly, $e^{\mathbf{A}t} = \mathbf{V}e^{\mathbf{S}t}\mathbf{V}^{-1}$ and
$e^{\mathbf{A}^{\text{T}}t}=(\mathbf{V}^{-1})^{\text{T}}e^{\mathbf{S}t}\mathbf{V}^{\text{T}}$. As a consequence, Eq.~\ref{finalH} is replaced by
$H_{ij}=\sum_{\alpha=1}^N\sum_{\beta=1}^N\frac{V_{i\alpha}(V^{-1})_{\alpha c}(V^{-1})_{\beta c}V_{j\beta}}
{\lambda_{\alpha} + \lambda_{\beta}}\left(1 - e^{-\left(\lambda_{\alpha} + \lambda_{\beta}\right)T_f}\right)$.
Therefore, the scaling laws in Eqs.~\ref{result} and \ref{upperresult} keep unchanged while the decaying exponents are replaced by
$2\text{Re}\lambda_N$ and $2\text{Re}\lambda_1$ respectively. Moreover, for large $T_f$ and PD $\mathbf{A}$, the constant in Eq.~\ref{result}
is still proportional to $2a_{cc}$ by using first-order approximation in~\cite{RWLL:PRL10}.

In conclusion, we have reduced the complexity of the fundamental problem of control cost
from the complicated and intractable Gramian matrix to the simple system matrix which is directly related to the network
structure. Our results have revealed that energy cost of controlling complex networks has different scaling behaviors with control time
in two time scales, separated by the characteristic time, $\frac{1}{2|\text{Re}\lambda_N|}$ and $\frac{1}{2|\text{Re}\lambda_1|}$
for the lower and the upper bound respectively. In the small-time regime, setting a relatively longer time for control always
leads to less energy cost. While, in the large-time regime, there exists the situation where we cannot reduce the
energy cost even given much more time. Furthermore, our results indicate that the lower (upper) bound of energy cost is less when controlling a randomized network with heterogeneous (homogeneous) node-degrees. These implications are important when considering the trade-off between
the energy cost and the control time, which may find applications not only for classical \cite{LSB:2011,controlRMP} but also for biological \cite{biologycontrolbook1,systembiology,RajaBiology} and quantum \cite{quantum_network} networks. Although we have given some heuristics,
a method to choose an optimal control node-set for minimizing the energy cost is lack, which is a promising future work.

We thank Drs. Maho Nakata and W.-X. Wang for helpful discussions.
YCL thanks the National University of Singapore
for great hospitality, and he is supported by AFOSR under
Grant No. FA9550-10-1-0083. GY and CHL are supported by DSTA of
Singapore under Grant No. POD0613356.

\newpage

\setcounter{figure}{0}
\setcounter{equation}{0}

\renewcommand\thefigure{S\arabic{figure}}
\renewcommand\theequation{S\arabic{equation}}
\renewcommand\bibnumfmt[1]{[S#1]}
\renewcommand{\citenumfont}[1]{S#1}

\begin{center}
\Large{\textbf{Supplemental Materials}}
\end{center}

\begin{center}
\textit{for ``Controlling complex networks: How much energy is needed?''}
\end{center}

\vspace{0.5cm}

\section{Decay behaviors of $E_{\text{min}}$ and $E_{\text{max}}$
in borderline cases}

\begin{figure}[!htp]
\begin{center}
\includegraphics[width=\columnwidth]{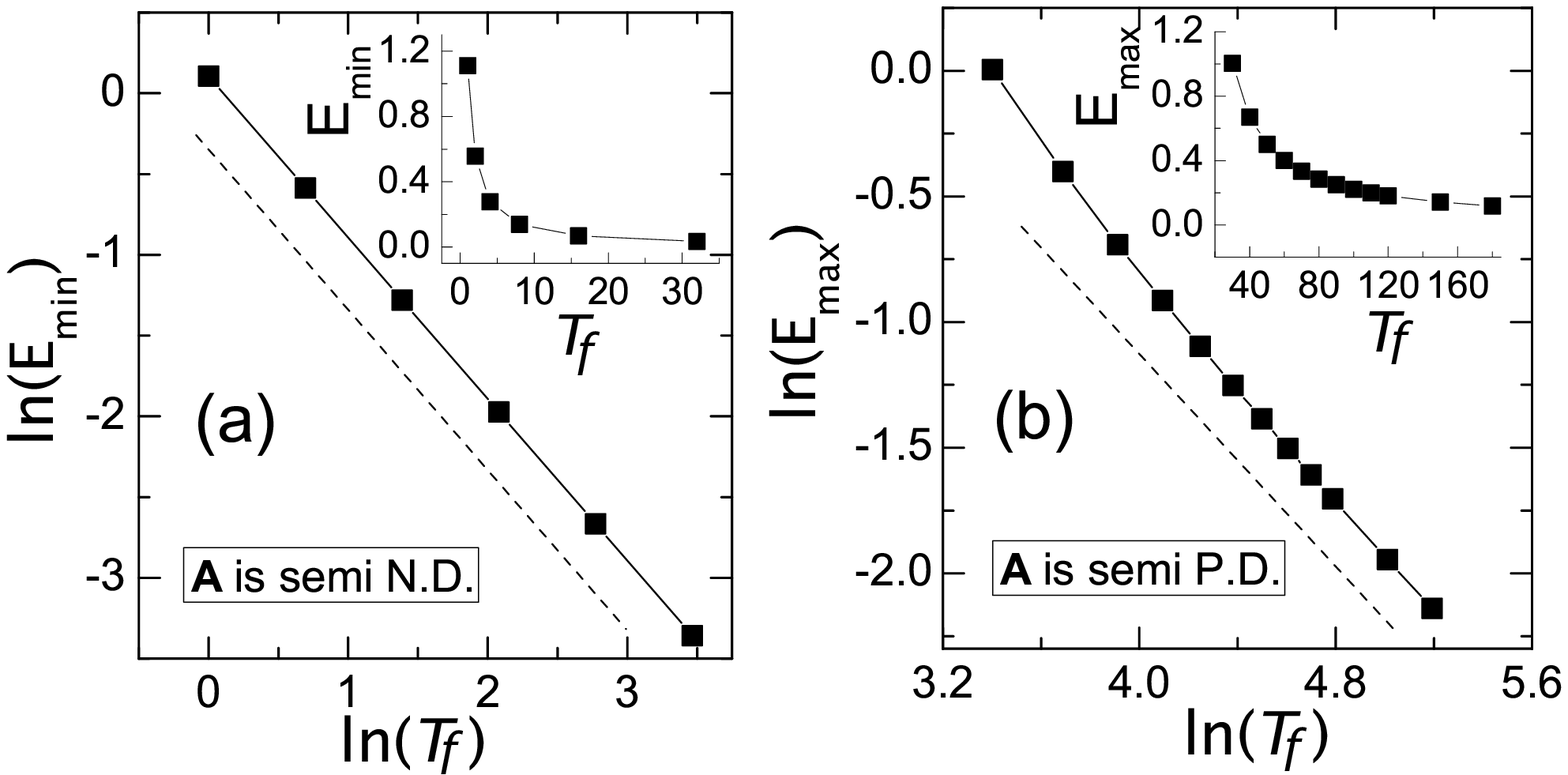}
\end{center}
\caption{Power-law decays of $E_{\text{min}}$ and $E_{\text{max}}$ for
the borderline situations. The dashed lines in (a) and (b) have the
same slope $-1$.}
\label{semi}
\end{figure}

In the main text we argue that, for large control time $T_f$, if
$\mathbf{A}$ is semi positive definite (PD), the lower bound of the
energy cost $E_{\text{min}}$ will decay as $T_f^{-1}$ and,
if $\mathbf{A}$ is semi negative definite (ND) the upper bound of
the energy cost $E_{\text{max}}$ will also decay as $T_f^{-1}$.
To provide numerical confirmation for these theoretical results,
we consider the situation of controlling an undirected network of 20 nodes
which can be controllable by one single driver node. Just as in the main text,
let nodal dynamics be $a_{ii} = -(a + s_i)$ where $s_i$ is the strength of node $i$.
Setting $a = 0$ so that $\mathbf{A}$ is semi ND, we obtain the decay
behavior of $E_{\text{min}}$, as depicted in Fig. \ref{semi}(a).
However, if $a = -14.148$, $\mathbf{A}$ becomes semi PD.
In this case, we observe the decay of $E_{\text{max}}$ as shown
in Fig. \ref{semi}(b). The dashed lines in both figures have the
same slope $-1$. Thus, in these borderline situations, $E_{\text{min}}$
and $E_{\text{max}}$ decay as $T_f^{-1}$ for relatively
large $T_f$, which is consistent with our theoretical predictions.

\section{Optimal control route}

\begin{figure}
\begin{center}
\includegraphics[width=\columnwidth]{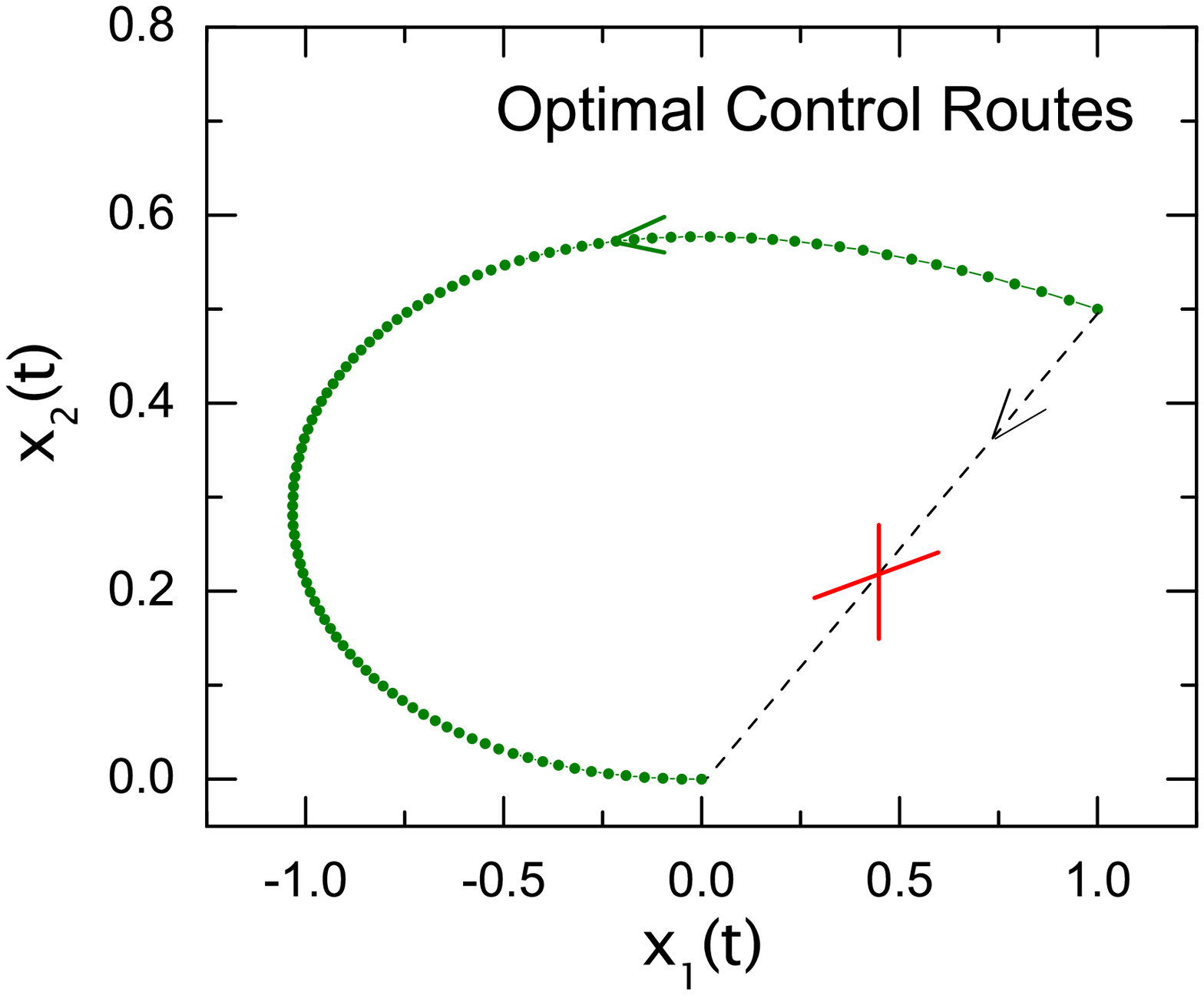}
\end{center}
\caption{(color online). Optimal control routes to steer a simple directed
network from the initial state $(1.0, 0.5)^\text{T}$ to the desired state $(0, 0)^\text{T}$
in the allowed time range [0,1].}
\label{routes}
\end{figure}

In the main text, we argue that the upper bound of the energy cost
associated with controlling a complex network can be very large,
because even the optimal control route of steering the whole network
from some initial state to the origin (zero state) is in general
highly circuitous, though smooth. Here we provide numerical result of
the optimal route for a simple directed network used in
\cite{MotterComment}:
$$
\mathbf{A} =
\begin{pmatrix}
1 & 0 \\
1 & 0
\end{pmatrix}, \hspace{0.5cm}
\mathbf{B} =
\begin{pmatrix}
1\\
0
\end{pmatrix}.
$$
Figure \ref{routes} shows that, when steering the network from
$\mathbf{x}_0 = (1.0, 0.5)^\text{T}$ to the desired state $\mathbf{x}_{T_f} = (0, 0)^\text{T}$
in the allowed time range $[0,1]$, the optimal route is indeed not
direct (dashed line) but smooth and circuitous ({\color{OliveGreen} $\bullet$}).
For undirected networks, we obtain similar highly circuitous optimal routes
(not shown here).

\section{Notes on structural equivalence of randomized networks}

As we stated in the main text, the constant value $\varepsilon(\mathbf{A},c)$ of energy cost in Eq. (8) is the inverse of
smallest eigenvalue of the correlation matrix $\mathbf{H}^{\infty}$ with elements $H^{\infty}_{ij} = \sum^N_{\alpha=1}\sum^N_{\beta}\frac{V_{i\alpha}V_{c\alpha}V_{c\beta}V_{j\beta}}{\lambda_{\alpha}+\lambda_{\beta}}$. Although it is
difficult, if not impossible, to estimate the smallest eigenvalue of $\mathbf{H}^{\infty}$, we can obtain some indications from the
viewpoint of correlations between nodes. When nodes are more structurally similar, they tend to have more similar dynamical
correlations with other nodes so that corresponding columns and rows in $\mathbf{H}^{\infty}$ become more similar. As a consequence,
the smallest eigenvalue will be less and $\mathbf{H}^{\infty}$ tend to be ill-conditioned so that the upper bound of control
energy cost increases drastically.

Because the energy cost of controlling a network depends not only on the network topology but also on the link weights and nodal dynamics, for
the convenience of comparing the impacts of network structures on the control cost, one may set the networks with equal size, edge numbers, but
different degree distributions, and set all nodal dynamics as a function of node-degrees. At this situation, when two nodes are \emph{structural equivalent},
the network cannot be controllable by any other driver node. Here \emph{structural equivalent} is defined for any two nodes $a$ and $b$ with the same
degree as follows: nodes $a$ and $b$ are structural equivalent if and only if there exists a nontrivial automorphism $f$ of the network~\cite{graphtheory} such that $f(a) = b$. More apparently, take unweighted undirected networks for example, denote the neighbor-set of any node $i$ by $\mathcal{S}(i)$, nodes $a$ and $b$ are structural equivalent if any of the following conditions are satisfied: \\
1. $\mathcal{S}(a) = \mathcal{S}(b)$, or \\
2. $a$ and $b$ are connected with each other and $(\mathcal{S}(a)-\{b\}) = (\mathcal{S}(b)-a)$, or \\
3. $\forall v_1\in \mathcal{S}(a)$, $\exists v_2\in\mathcal{S}(b)$ such that $(\mathcal{S}(v_1)-a)=(\mathcal{S}(v2)-b)$ or $v_1$ and
$v_2$ are connected with each other and $(\mathcal{S}(v_1)-a-v_2)=(\mathcal{S}(v_2)-b-v_1)$, or  \\
4. higher-order neighbors of $a$ and $b$ meet condition 3.

Therefore, for randomized networks, the probability of
that any two nodes are structural equivalent can provide hints for the energy cost of controlling them: the larger the probability,
the more the energy needed for controlling. We here focus on the probabilities of the most possible conditions 1 and 2, and ignore the high-order
probabilities of conditions 3 and 4.

We will derive the probability of that two nodes are structural equivalent for a randomized network with degree distribution $p(k)$.
For a randomized network, two nodes $i$ and $j$ are connected with the probability
$p_{ij} = \frac{k_ik_j}{\langle k\rangle N}$, where $k_i$ and $k_j$
are the degrees of nodes $i$ and $j$ respectively, $\langle k\rangle$ is the average degree and $N$ is the network size. Thus the probability of
two nodes $a$ and $b$ with the same degree $k_0$ have the same neighbors is
\begin{equation}
\begin{aligned}
p_{\text{eq}} &= \binom{k_0}{N-2}\prod^{k_0}_{i=1}p_{ai}p_{bi}\prod^{N-2}_{i=k_0+1}(1-p_{ai})(1-p_{bi})\\
              &= \binom{k_0}{N-2}\prod^{k_0}_{i=1}\bigg{(}\frac{k_0k_i}{\langle k\rangle N}\bigg{)}^2
                 \prod^{N-2}_{i=k_0+1}\bigg{(}1-\frac{k_0k_i}{\langle k\rangle N}\bigg{)}^2.
\end{aligned}
\nonumber
\end{equation}
As the degree $k_i$ of node $i$ is generated from the distribution $p(k)$ independently,
the expectation value of $p_{\text{eq}}$ can be obtained as $\langle p_{\text{eq}}\rangle
= \int_{k_1}\int_{k_2}\ldots\int_{k_{N-2}}[p_{\text{eq}}p(k_1)p(k_2)\ldots p(k_{N-2})]dk_1dk_2\ldots dk_{N-2}$
where the integrations begin at the smallest degree $k_\text{min}$ and end at the largest degree $k_\text{max}$.
Therefore, we have
\begin{equation}
\begin{aligned}
\langle p_{\text{eq}}\rangle = &Z\prod^{k_0}_{i=1}\int_{k_i}\bigg{(}\frac{k_i}{\langle k\rangle}\bigg{)}^2p(k_i)dk_i
                                 \prod_{i=k_0+1}^{N-2}\int_{k_i}\bigg{(}\frac{N}{k_0}- \\
                                 & \frac{k_i}{\langle k\rangle}\bigg{)}^2p(k_i)dk_i \\
                             = &Z\bigg{(}\frac{\langle k^2\rangle}{\langle k\rangle^2}\bigg{)}^{k_0}\bigg{(} (\frac{N}{k_0})^2 - 2(\frac{N}{k_0})
                                 + \frac{\langle k^2\rangle}{\langle k\rangle^2}\bigg{)}^{N-2-k_0},
\end{aligned}
\nonumber
\end{equation}
where $Z = \binom{k_0}{N-2}(k_0/N)^{2(N-2)}$. For the condition 2 mentioned above, one can obtain similar result. Thus, when $k_0 < N/2$, the larger the factor $\frac{\langle k^2\rangle}{\langle k\rangle^2}$,
the larger the expectation value $\langle p_{\text{eq}}\rangle$.

From the result one can expect that controlling randomized Barab\'{a}si-Albert networks need much more energy than
controlling Erd\"{o}s-R\'{e}nyi networks, as shown in Fig.~3(a) in the main text. Moreover, for randomized scale-free networks with degree distribution $p(k) \propto k^{-\gamma}$ ($\gamma > 2$),
the smaller the value of $\gamma$, the more the energy needed for controlling. Especially, if ignore the nodal dynamics, when $\gamma \rightarrow 2$ the
value of $p_{\text{eq}} \rightarrow 1$ for most of degrees $k_0$ because $\frac{\langle k^2\rangle}{\langle k\rangle^2}$ is very large at this situation. The ending result is that controlling a scale-free network with $\gamma = 2$ needs an infinite amount of energy. In other words, the network is uncontrollable
 at $\gamma = 2$ unless most of the nodes are set as driver nodes. This indication is qualitatively consistent with the result found in~\cite{LSB:2011nature}.

\section{Reachability: from initial state $\mathbf{x}_0 = 0$ to desired
state $\mathbf{x}_{T_f} \neq 0$} \label{sec:reachability}

In the main text, we consider the case of controlling a networked system
from an arbitrary state $\mathbf{x}_0 \neq 0$ to the origin $\mathbf{x}_{T_f} = 0$,
which defines \emph{controllability}. Here, we consider the case of steering
the system described in Eq. (1) from $\mathbf{x}_0 = 0$ to
$\mathbf{x}_{T_f} \neq 0$. This situation is referred to as
\emph{reachability} \cite{linearcontrol}.

To analyze the energy cost associated with reachability, we write down
the energy expression in the complete form as
$\mathcal{E}(T_f) = (\mathbf{x}_{T_f}^\text{T} - \mathbf{x}_0^\text{T}e^{\mathbf{A}^\text{T}T_f})
\mathbf{W}_{T_f}^{-1}(\mathbf{x}_{T_f} - e^{\mathbf{A}T_f}\mathbf{x}_0)$,
where $\mathbf{W}_{T_f} =
\int_0^{T_f}e^{\mathbf{A}t}\mathbf{BB}^\text{T}e^{\mathbf{A}^\text{T}t}dt$.
Since $\mathbf{x}_0 = 0$, we have
$\mathcal{E}(T_f) = \mathbf{x}_{T_f}^\text{T}\mathbf{W}_{T_f}^{-1}\mathbf{x}_{T_f}$.
For undirected networks, we factorize $\mathbf{A}$ as $\mathbf{A} = \mathbf{VSV}^{\text{T}}$,
where $\mathbf{V}$ is the orthonormal eigenvector matrix with
$\mathbf{VV}^{\text{T}}=\mathbf{V}^{\text{T}}\mathbf{V}=\mathbf{I}$
and $\mathbf{S} = \text{diag}\{\lambda_1, \lambda_2, \ldots, \lambda_N\}$
with descending order: $\lambda_1>\lambda_2>\ldots>\lambda_N$.
Assume that only the $c$-th node is controlled, we have
\begin{equation} \label{finalW}
W_{ij}=\sum_{\alpha=1}^N\sum_{\beta=1}^N\frac{V_{i\alpha}V_{c\alpha}
V_{c\beta}V_{j\beta}}{\lambda_{\alpha} + \lambda_{\beta}}(e^{(\lambda_{\alpha} +
\lambda_{\beta})T_f} - 1).
\end{equation}
Note that the term in the parenthesis
is different from that of Eq. (6) in the main text.

\begin{figure}
\begin{center}
\includegraphics[width=0.7\columnwidth]{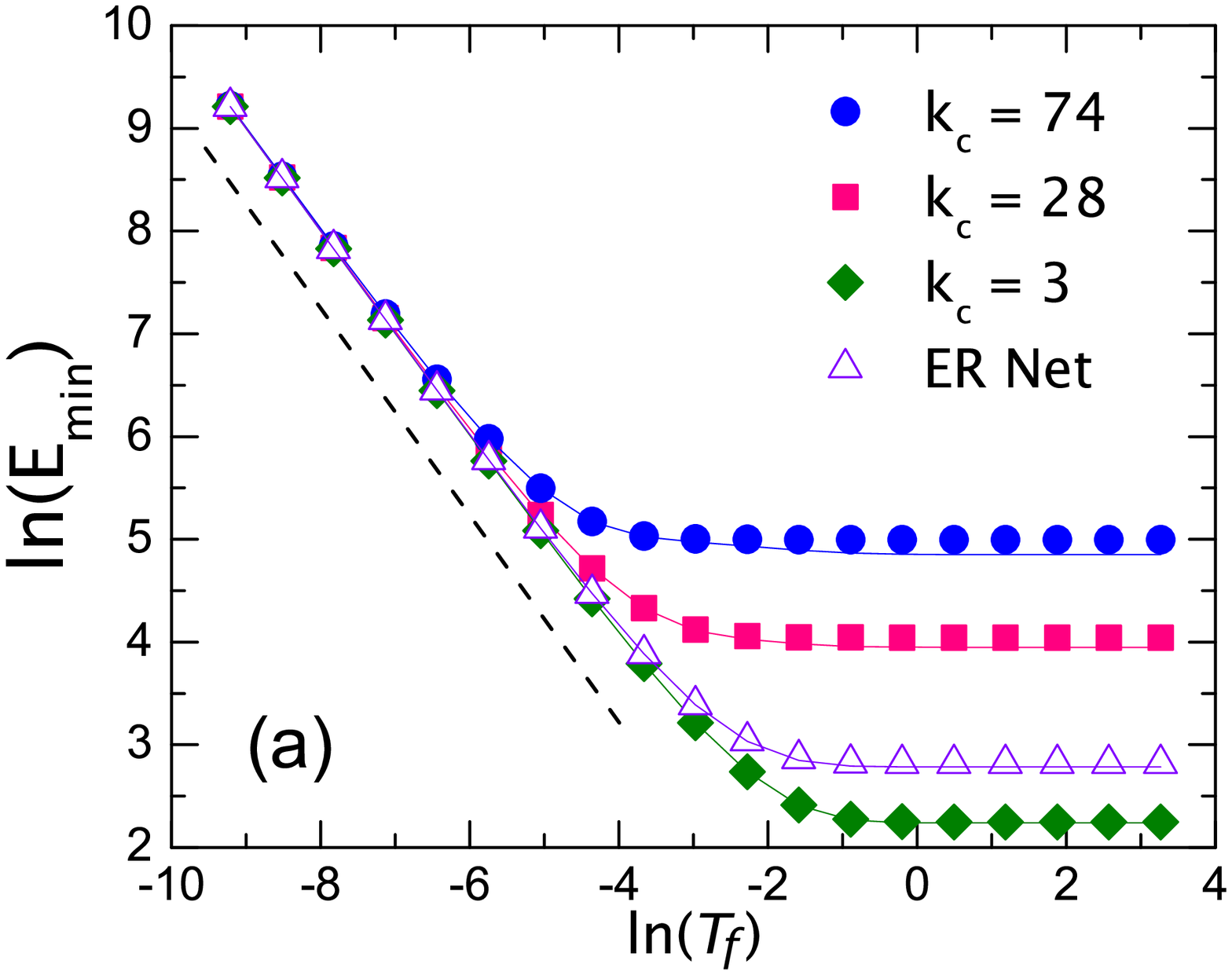}
\includegraphics[width=0.85\columnwidth]{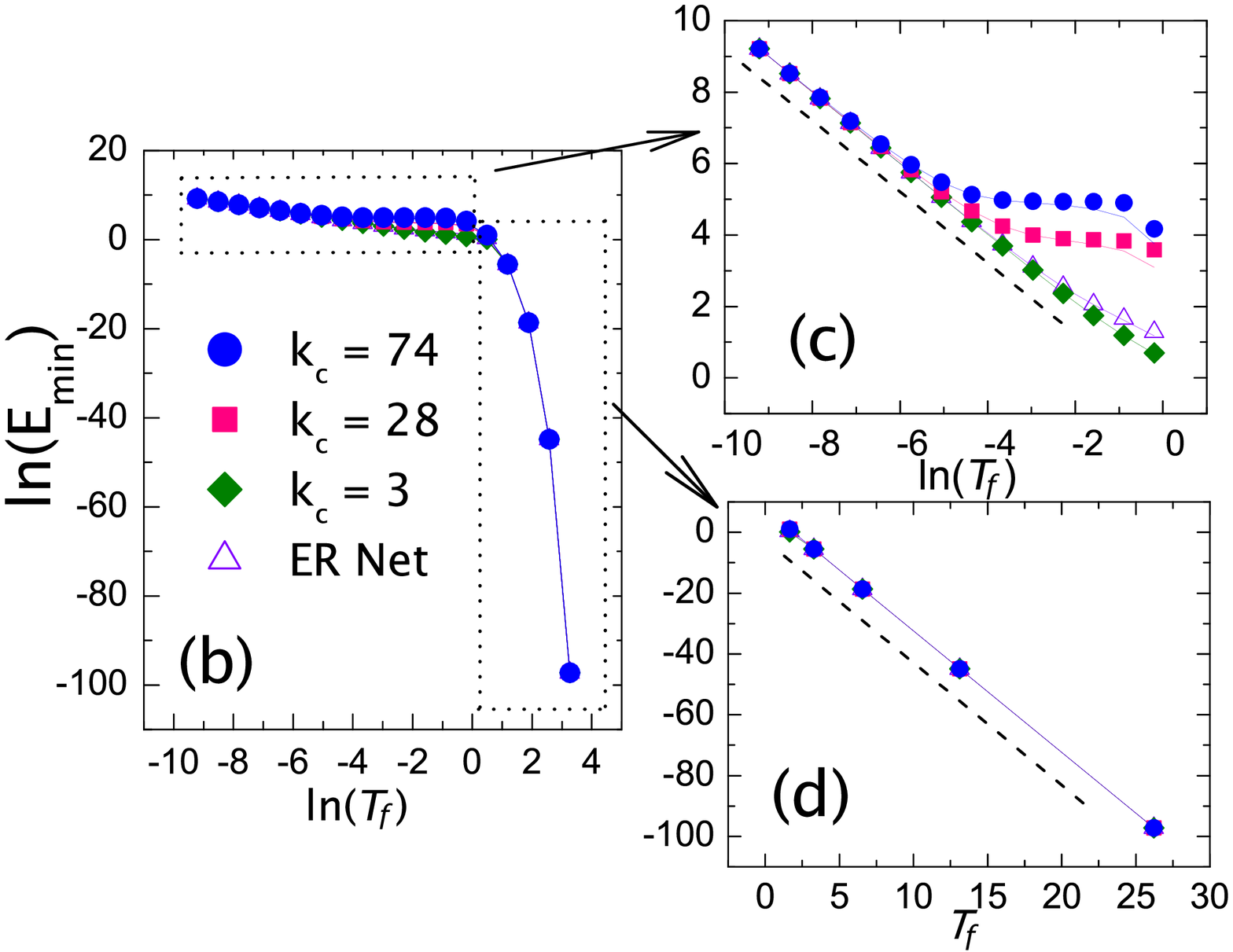}
\end{center}
\caption{(color online). In the context of reachability where a network
is controlled from $\mathbf{x}_0 = 0$ to the desired state
$\mathbf{x}_{T_f} \neq 0$, the lower bound of the
energy cost $E_{\text{min}} = 1/\xi_{\text{max}}$ versus the control time $T_f$.
All networks considered are scale-free except the one noted as ER Net in (a),
which is an Erd\"os-R\'eyni random network. The networks have the same size
$N = 500$ and the same average degree $\langle k \rangle = 6$. In all figures,
$k_c$ is the degree of the directly controlled node, and the symbols represent
the results of our numerical computation and the corresponding solid lines are
the results of our estimation $\xi_{\text{max}} \approx \text{Tr}[\mathbf{W}]$.
In (a), the parameter is $a = 2$, which makes the matrix $\mathbf{A}$ ND. In (b-d),
$a = -2$ so that $\mathbf{A}$ is not ND. The dashed lines in the log-log plots in
(a) and (c) have the same slope $-1$, which agrees with our theoretical result
$T_f^{-1}$ for small $T_f$. The dashed line in the semi-log plot in (d) has the
slope $-3.99\pm0.01$, which corresponds to our theoretical estimate
$e^{-2\lambda_{1}T_f}$, as $-2\lambda_{1} = 2a = -4.0$ in
this case (see the text for details).}
\label{SEvsT}
\end{figure}

Recall the normalized energy cost
$E(T_f) = \mathcal{E}(T_f)/\parallel \mathbf{x}_{T_f}\parallel^2$,
which satisfies
\begin{equation} \label{Sbound}
1/\xi_{\text{max}} = E_{\text{min}} \leq E(T_f) \leq  E_{\text{max}} = 1/\xi_{\text{min}},
\end{equation}
where $\xi_{\text{min}}$ and $\xi_{\text{max}}$ are the minimal and maximal
eigenvalues of the matrix $\mathbf{W}_{T_f}$, respectively.
Following a similar analysis in the main text, we approximate
$\xi_{\text{max}}$ by the trace of $\mathbf{W}_{T_f}$ (verified again
numerically, as shown in Fig. \ref{SEvsT}) and get that
\begin{equation}
\begin{aligned}
\xi_{\text{max}} & \approx \sum_{i = 1}^{N}W_{ii} \ \ \ [\equiv   \text{Tr}(\mathbf{W}_{T_f})] \\
                        & = \sum_{i}^{N}\sum_{\alpha=1}^N\sum_{\beta=1}^N\frac{V_{i\alpha}V_{c\alpha}V_{c\beta}V_{i\beta}}{\lambda_{\alpha} + \lambda_{\beta}}(e^{(\lambda_{\alpha} + \lambda_{\beta})T_f}-1)\\
                        & = \sum_{\alpha=1}^N\sum_{\beta=1}^N\frac{V_{c\alpha}V_{c\beta}}{\lambda_{\alpha} + \lambda_{\beta}}(e^{(\lambda_{\alpha} + \lambda_{\beta})T_f}-1)(\sum_{i=1}^{N}V_{i\alpha}V_{i\beta}) \\
                        & = \sum_{\alpha=1}^N\sum_{\beta=1}^N\frac{V_{c\alpha}V_{c\beta}}{\lambda_{\alpha} + \lambda_{\beta}}(e^{(\lambda_{\alpha} + \lambda_{\beta})T_f}-1)\delta_{\alpha\beta} \\
                        & = \sum_{\alpha=1}^N\frac{V_{c\alpha}V_{c\alpha}}{2\lambda_{\alpha}}(e^{2\lambda_{\alpha}T_f}-1).
\label{SEQlower}
\end{aligned}
\end{equation}

\begin{figure}[!htp]
\begin{center}
\includegraphics[width=0.49\columnwidth]{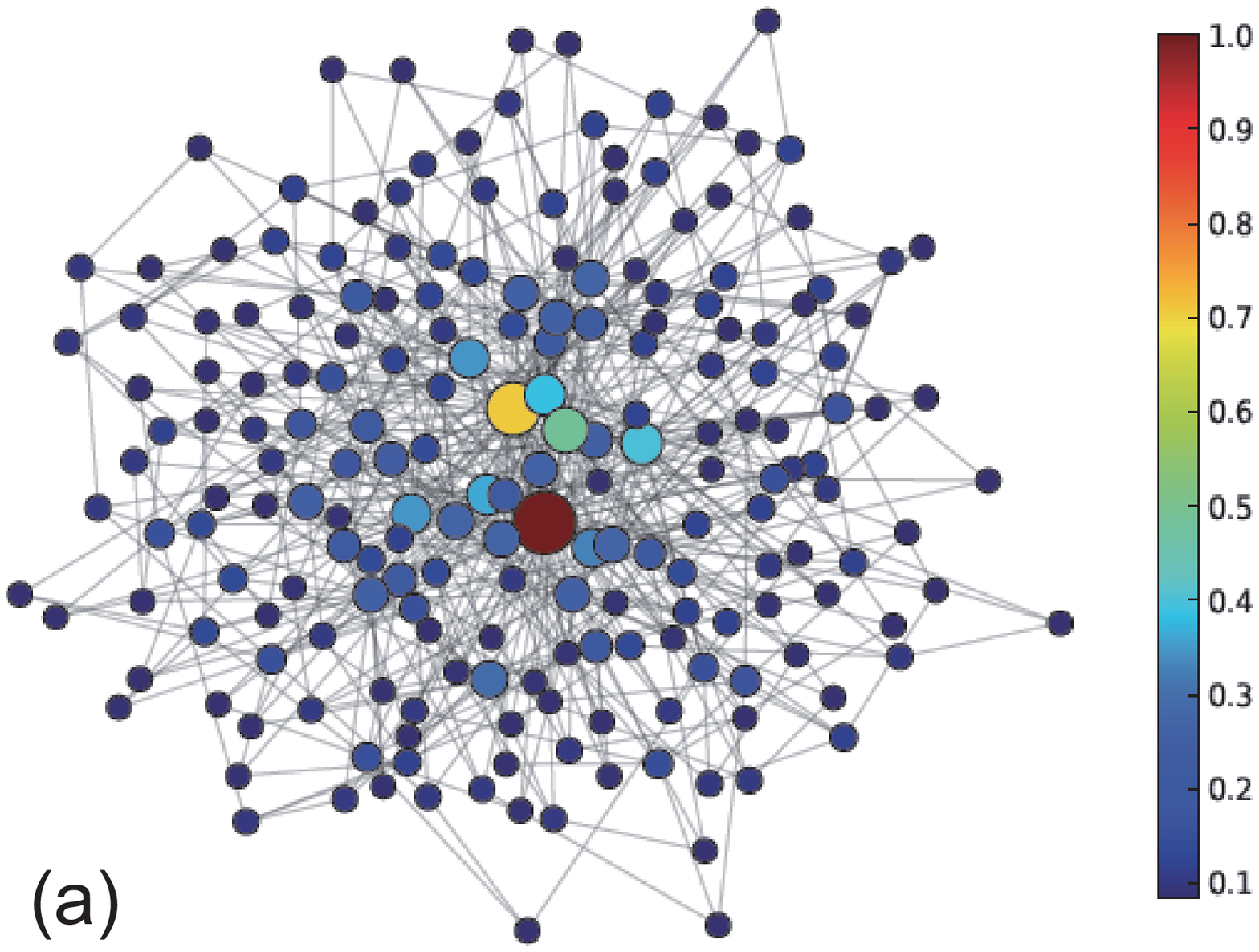}
\includegraphics[width=0.49\columnwidth]{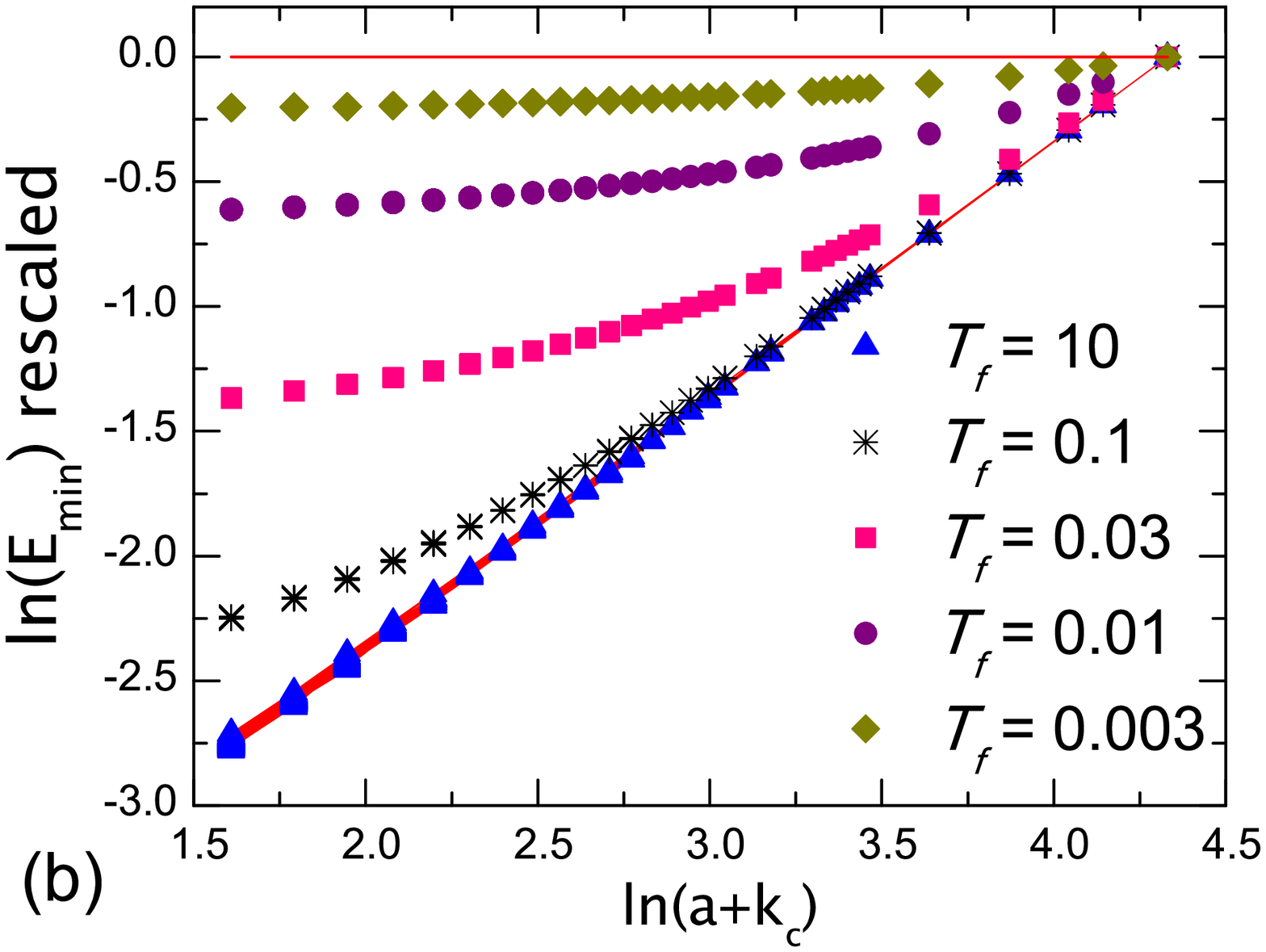}
\includegraphics[width=0.49\columnwidth]{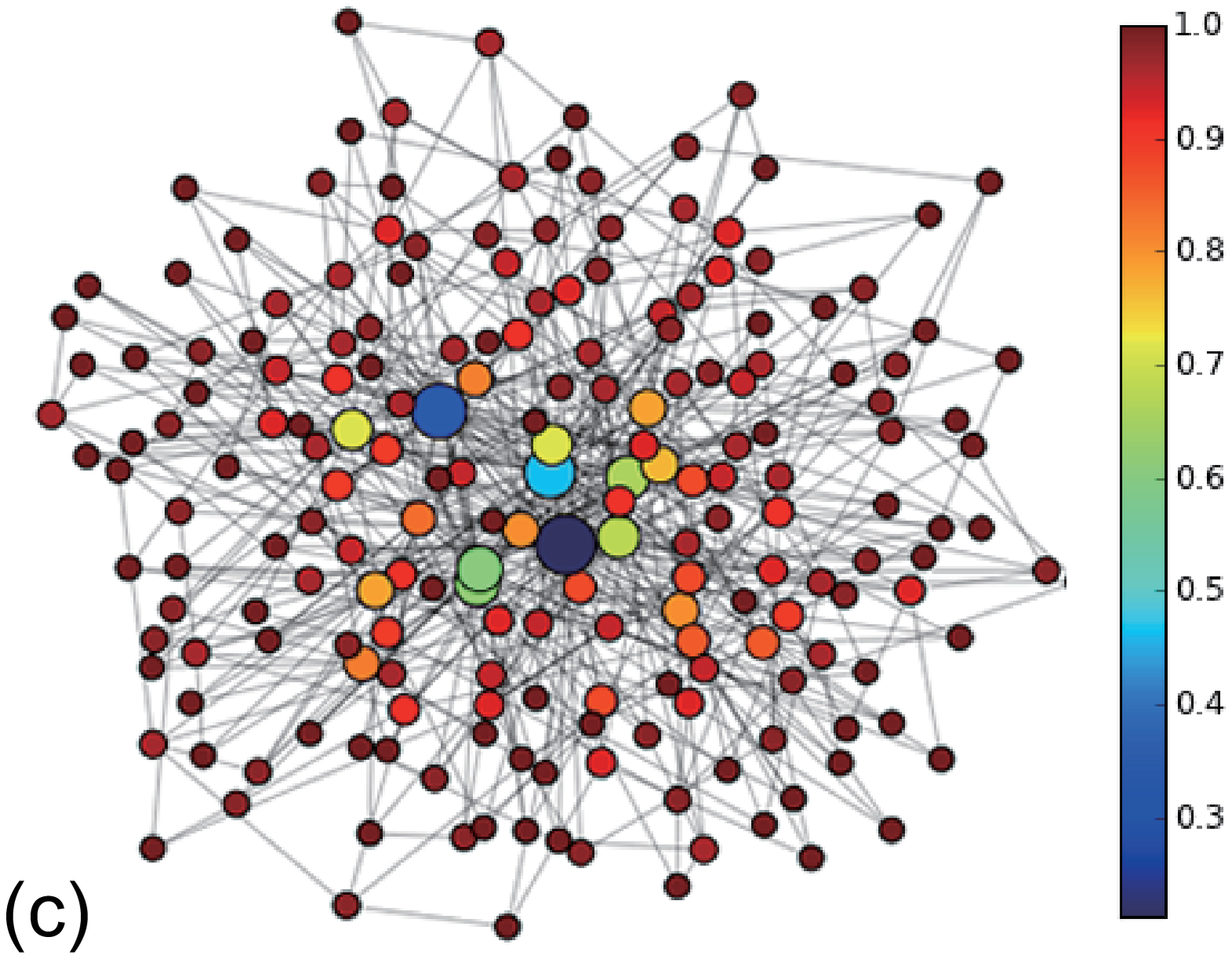}
\includegraphics[width=0.49\columnwidth]{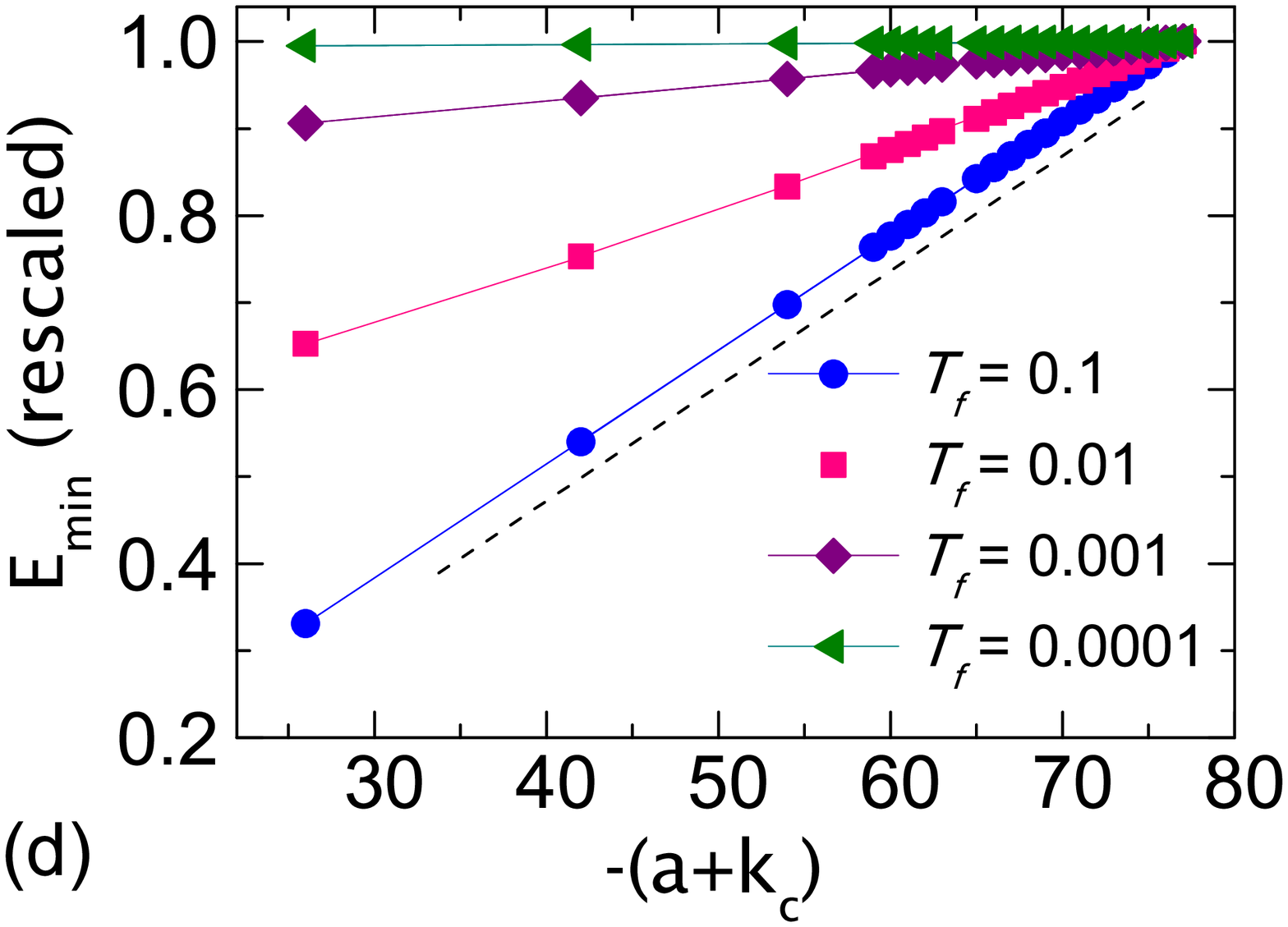}
\end{center}
\caption{(color online). (a,b) For the context of reachability, the
lower bound of the energy cost $E_{\text{min}} = 1/\xi_{\text{max}}$ versus
the degree $k_c$ of the directly controlled node. The scale-free networks have
the same size $N = 200$ and average degree $\langle k\rangle = 6$,
and the largest degree is $k_m = 54$. For comparison, (c) and (d)
show the corresponding results for the case of \emph{controllability}
as discussed in the main text, i.e., steering the network from
$\mathbf{x}_0 \neq 0$ to the desired state $\mathbf{x}_{T_f} = 0$.
In (a) and (b), we set $a = 2$ so that $\mathbf{A}$ is ND and, hence,
following the analysis in Sec. \ref{sec:reachability}, we find that
$E_{\text{min}}$ converges to a constant value. In (c) and (d), we set
$a = -80$ so that $\mathbf{A}$ becomes PD and, hence, following the
analysis in the main text, $E_{\text{min}}$ converges to a constant value.
While (b) and (d) display the relationship between the lower bound of the
energy cost and the degree $k_c$ of the controlled single node for
different values of the allowed control time $T_f$, (a) and (c) show the behaviors
for large time $T_f = 10$, where the node size represents its degree, and the
node color represents $E_{\text{min}}$ when controlling the corresponding
node only. Note that the bottom curves in both (b) and (d) show
the behavior $E_{\text{min}} \propto |a + k_c|$, which are consistent with
our theoretical result
$E_{\text{min}} \approx |1/(\mathbf{A} + \mathbf{A}^\text{T})^{-1}| \approx 2|a+k_c|$.}
\label{SEvsKc}
\end{figure}

From Eq. (\ref{SEQlower}), we obtain the behavior of the lower bound
$E_{\text{min}} = 1/\xi_{\text{max}}$:\\
\emph{Small $T_f$ regime}: $1/\xi_{\text{max}} \approx 1/\sum_{\alpha=1}^N\frac{V_{c\alpha}V_{c\alpha}}{2\lambda_{\alpha}}
(e^{2\lambda_{\alpha}T_f}-1) \approx 1/\sum_{\alpha=1}^NV_{c\alpha}V_{c\alpha}T_f \approx 1/T_f$.\\
\emph{Large $T_c$ regime}: if $\mathbf{A}$ is ND,
i.e., all of $\mathbf{A}$'s eigenvalues are negative, we have
$1/\xi_{\text{max}} \approx -1/\sum_{\alpha=1}^N\frac{V_{c\alpha}V_{c\alpha}}
{2\lambda_{\alpha}} = -1/[(\mathbf{A} + \mathbf{A}^\text{T})^{-1}]_{cc}$.
If $A$ is not ND, i.e., at least one of $A$'s eigenvalues is positive,
$E_{\text{min}}$ will vanish exponentially as
$1/\xi_{\text{max}} \sim e^{-2\lambda_1T_f}$,
where $\lambda_1$ is the most positive eigenvalue of $\mathbf{A}$.
For the borderline situation that $\mathbf{A}$ is semi-ND, we can obtain the
large-time decay behavior by setting $\lambda_1 \rightarrow 0$ in
Eq. (\ref{SEQlower}), which gives $E_{\text{min}} \sim 1/T_f$.

The behaviors of the lower bound of the energy cost associated with the
\emph{reachability} can then be summarized, as follows.
\begin{equation}
E_{\text{min}}
  \begin{cases}
     \approx T_f^{-1} & \text{small $T_f$}\\
     \approx -\frac{1}{[(\mathbf{A}+\mathbf{A}^\text{T})^{-1}]_{cc}} & \text{large $T_f$, $\mathbf{A}$ is ND}\\
     \xrightarrow[\sim \exp{\left(-2\lambda_1T_f\right)}]{\sim~T_f^{-1}}0  & \text{large $T_f$, $\mathbf{A}$ is } \frac{\text{semi ND}}{\text{not ND}}
  \end{cases}
\label{SEQlowerresult}
\end{equation}

We take unweighted networks for example, and show the numerical results on
the behaviors of $E_{\text{min}}$ versus the allowed control
time $T_f$ in Fig. \ref{SEvsT}. We have also studied the relationship
between $E_{\text{min}}$ and the degree $k_c$ of the directly controlled node,
as shown in Fig. \ref{SEvsKc}. We see that, if the matrix $\mathbf{A}$ is ND,
$E_{\text{min}}$ associated with \emph{reachability} will converge to a
constant value. In contrast, if $\mathbf{A}$ is PD, $E_{\text{min}}$ associated
with \emph{controllability} will converge to a constant value. Furthermore, we
recall that, in the context of controllability in the main text, driving a node
with higher degree will induce smaller $E_{\text{min}}$. However, in the context
of reachability treated here, driving a node with a higher degree will induce
larger $E_{\text{min}}$, as shown in Figs. \ref{SEvsKc}(a) and \ref{SEvsKc}(c).

Following a similar analysis in the main text, we obtain the behaviors of the
energy upper bound $E_{\text{max}} = 1/\xi_{min}$ associated with \emph{reachability},
as follows (numerical results not shown here).
\begin{equation} \label{Supperresult}
 E_{\text{max}}
  \begin{cases}
     \approx T_f^{-\nu} (\nu \gg 1) & \text{small $T_f$}\\
     \approx \varsigma(\mathbf{A},c) & \text{large $T_f$, $\mathbf{A}$ is not PD}\\
     \xrightarrow[\sim \exp{\left(-2\lambda_NT_f\right)}]{\sim~T_f^{-1}}0 & \text{large $T_f$, $\mathbf{A}$ is } \frac{\text{semi PD}}{\text{PD}},
  \end{cases}
\end{equation}
where $\varsigma(\mathbf{A},c)$ is a positive value depending on $\mathbf{A}$
and $c$, and $\lambda_N$ is the eigenvalue of $\mathbf{A}$ with the least positive
real part.

\section{Derivation of optimal control $\mathbf{u}_t$}

In this paper we have studied the behaviors of the energy cost
$\mathcal{E}(T_f) = \int_0^{T_f}\mathbf{u}_{t}^\text{T}\mathbf{u}_{t}dt$
associated with controlling complex networks while choosing
$\mathbf{u}_{t} = \mathbf{B}^\text{T}e^{\mathbf{A}^\text{T}(T_f-t)}\mathbf{W}_{T_f}^{-1}\mathbf{v}_{T_f}$,
where $\mathbf{W}_{T_f} = \int_0^{T_f}e^{\mathbf{A}t}\mathbf{B}\mathbf{B}^\text{T}e^{\mathbf{A}^\text{T}t}dt$,
and $\mathbf{v}_{T_f}=\mathbf{x}_{T_f} - e^{\mathbf{A}T_f}\mathbf{x}_0$. We state in the main text
without proof that this form of $\mathbf{u}_{t}$ minimizes the energy cost $\mathcal{E}(T_f)$.
Although the derivation of this statement can be found in books on \emph{optimal control}
(e.g., \cite{Soptimal}, among others) and is not the subject of our present paper, we would like to
include it here for readers' convenience.

\noindent\emph{System}:
$\mathbf{\dot{x}}_{t} = \mathbf{Ax}_{t} + \mathbf{B\tilde{u}}_{t}$,
$\mathbf{x}(0) = \mathbf{x}_0$, $\mathbf{x}(T_f) = \mathbf{x}_{T_f}$.\\
\emph{Problem}: Choose an optimal $
\mathbf{u}_t$ out of $\tilde{\mathbf{u}}_{t}$: $[0,T_f] \rightarrow \mathbf{R}^N$
to minimize $J = \int_0^{T_f}\mathbf{\tilde{u}}_{t}^\text{T}\mathbf{\tilde{u}}_{t}dt$.

The optimal problem can be solved by using Pontryagin's Maximum Principle (PMP).
Firstly, define the Hamiltonian
\begin{equation}
M_{t} = \mathbf{\tilde{u}}_{t}^\text{T}\mathbf{\tilde{u}}_{t} + \mathbf{\lambda}_{t}^\text{T}
(\mathbf{Ax}_{t} + \mathbf{B{\tilde u}}_t)
\end{equation}
where $\mathbf{\lambda}_{t}$ is the vector of Lagrange multipliers. Then,
according to the PMP, the optimal control signal $\mathbf{u}_{t}$ obeys
\begin{subequations}
\begin{eqnarray}
\mathbf{\dot{x}}_{t} = \Big(\frac{\partial M_{t}}{\partial \mathbf{\lambda}_{t}}\Big)^\text{T}\Big |_{\mathbf{u}_{t}} &=& \mathbf{Ax}_{t} + \mathbf{Bu}_{t},\\
-\mathbf{\dot{\lambda}}_{t} = \Big(\frac{\partial M_{t}}{\partial \mathbf{x}_{t}}\Big)^\text{T}\Big|_{\mathbf{u}_{t}} &=& \mathbf{A}^\text{T}\mathbf{\lambda}_{t},\\
0 = \Big(\frac{\partial M_{t}}{\partial \mathbf{\tilde{u}}_{t}}\Big)^\text{T}\Big|_{\mathbf{u}_{t}} &=& \mathbf{u}_{t} + \mathbf{B}^\text{T}\mathbf{\lambda}_{t}.
\end{eqnarray}
\end{subequations}
The solution of Eq. (S7b) is
$\lambda_{t} = e^{-\mathbf{A}^\text{T}t}\mathbf{c}$,
where $\mathbf{c}$ is a vector independent of time $t$.
Substituting it into Eq. (S7c), we get
\begin{equation}
\mathbf{u}_{t} = -\mathbf{B}^\text{T}e^{-\mathbf{A}^\text{T}t}\mathbf{c}.
\end{equation}
The solution of Eq. (S7a) under the boundary condition
$\mathbf{x}_{t = 0} = \mathbf{x}_0$ is
\begin{equation}
\mathbf{x}_{t} = e^{\mathbf{A}t}(\mathbf{x}_0 + \int_0^{t}e^{-\mathbf{A}s}\mathbf{Bu}_{s}ds).
\end{equation}
Substituting Eq. (S8) into Eq. (S9) gives
$\mathbf{x}_{t} = e^{\mathbf{A}t}\mathbf{x}_0 - (\int_0^{t}e^{\mathbf{A}(t - s)}
\mathbf{B}\mathbf{B}^\text{T}e^{\mathbf{A}^\text{T}(t - s)}ds)e^{-\mathbf{A}^\text{T}t}\mathbf{c}$.
By a change of variables $t - s = \tau$ and thus $d\tau = -ds$ in the integral, we
can rewrite the equation as
$\mathbf{x}_{t} = e^{\mathbf{A}t}\mathbf{x}_0 - (\int_0^{t}e^{\mathbf{A}\tau}
\mathbf{B}\mathbf{B}^\text{T}e^{\mathbf{A}^\text{T}\tau}d\tau)e^{-\mathbf{A}^\text{T}t}\mathbf{c}$. Denoting
\begin{equation}
\mathbf{W}_{t} = \int_0^{t}e^{\mathbf{A}\tau}\mathbf{B}\mathbf{B}^\text{T}e^{\mathbf{A}^\text{T}\tau}d\tau
\end{equation}
and using the boundary condition $\mathbf{x}_{t = T_f} = \mathbf{x}_{T_f}$, we have
\begin{equation}
\mathbf{x}_{T_f} = e^{\mathbf{A}T_f}\mathbf{x}_0 - \mathbf{W}_{T_f}e^{-\mathbf{A}^\text{T}T_f}\mathbf{c},
\end{equation}
where $\mathbf{W}_{T_f}$ is symmetric as well as PD if the system is controllable. It is
also referred to as the Gramian matrix of controllability.
Thus
\begin{equation}
\mathbf{c} = -e^{\mathbf{A}^\text{T}T_f}\mathbf{W}_{T_f}^{-1}\mathbf{v}_{T_f},
\end{equation}
where
\begin{equation}
\mathbf{v}_{T_f} = \mathbf{x}_{T_f} - e^{\mathbf{A}T_f}\mathbf{x}_0.
\end{equation}
Substituting Eq. (S12) into Eqs. (S8) and (S9), we obtain
\begin{subequations}
\begin{eqnarray}
\mathbf{u}_{t} &=& \mathbf{B}^\text{T}e^{\mathbf{A}^\text{T}(T_f - t)}\mathbf{W}_{T_f}^{-1}\mathbf{v}_{T_f},\\
\mathbf{x}_{t} &=& e^{\mathbf{A}t}\mathbf{x}_0 + \mathbf{W}_{t}e^{\mathbf{A}^\text{T}(T_f - t)}\mathbf{W}_{T_f}^{-1}\mathbf{v}_{T_f},\\
\mathcal{E}(T_f) &=& J_{\text{optimal}} = \mathbf{v}_{T_f}^\text{T}\mathbf{W}_{T_f}^{-1}\mathbf{v}_{T_f}.
\end{eqnarray}
\end{subequations}


\begin{thebibliography}{99}

\bibitem{Barabasi:review}
A.-L. Barab\'asi, Science \textbf{325}, 412 (2009); Nature Physics \textbf{8}, 14 (2012).

\bibitem{Newman:book}
M. E. J. Newman, {\it Networks: An Introduction} (Oxford University Press, Oxford, UK, 2010).

\bibitem{DynamicalBook}
A. Barrat, M. Barthel\'emy, and A. Vespignani, {\it Dynamical Processes on Complex Networks} (Cambridge University Press, NY, USA, 2008).

\bibitem{PRL1997}
R. O. Grigoriev, M. C. Cross, and H. G. Schuster, Phys. Rev. Lett. \textbf{79}, 2795 (1997).

\bibitem{controlRMP}
J. Bechhoefer, Rev. Mod. Phys. \textbf{77}, 783 (2005).

\bibitem{LH:2007}
A. Lombardi and M. H\"ornquist, Phys. Rev. E \textbf{75}, 056110 (2007).

\bibitem{PhysicaD2010}
M. Porfiri and F. Fiorilli, Physica D \textbf{239}, 454 (2010).

\bibitem{LSB:2011}
Y.-Y. Liu, J.-J. Slotine and A.-L. Barab\'asi, Nature (London),
473, {\bf 167} (2011).

\bibitem{CowanarXiv}
N. J. Cowan, E. J. Chastain, D. A. Vilhena, J. S. Freudenberg, and C. T. Bergstrom, arXiv:1106.2573 (2011).

\bibitem{WangarXiv}
W.-X. Wang, X. Ni, Y.-C. Lai, and C. Grebogi, Phys. Rev. E \textbf{85}, 026115 (2012).

\bibitem{biologycontrolbook1}
I. Shmulevich, E. R. Dougherty, \emph{Probabilistic Boolean Networks: The Modeling and Control of
Gene Regulatory Networks} (SIAM, PA, USA, 2009).

\bibitem{RajaBiology}
I. Rajapakse, M. Groudine, and M. Mesbahi, Proc. Natl. Acad. Sci. U.S.A. \textbf{108}, 17257 (2011).

\bibitem{networkcontrolbook2}
M. Mesbahi and M. Egerstedt, \emph{Graph Theoretic Methods in Multiagent Networks} (Princeton University Press, NJ, USA, 2010).

\bibitem{Network_Control}
X. Li, X. F. Wang and G. Chen, IEEE Trans. Circ. Syst.-I: {\bf 51}, 2074 (2004).

\bibitem{LCWX:2008}
B. Liu, T. Chu, L. Wang and G. Xie, IEEE Trans. Automat. Contr.
{\bf 53}, 1009 (2008).

\bibitem{RJME:2009}
A. Rahmani, M. Ji, M. Mesbahi and M. Egerstedt,  SIAM J. Contr.
Optim. {\bf 48}, 162 (2009).

\bibitem{supplement}
Supplemental Materials associated with this paper.

\bibitem{controlbook1}
W. J. Rugh, \emph{Linear System Theory (2nd ed.)} (Prentice-Hall, NJ, USA, 1996).

\bibitem{systembiology}
E. Klipp, W. Liebermeister, C. Wierling, A. Kowald, H. Lehrach, and R. Herwig,
\emph{Systems Biology: A Textbook} (Wiley-VCH, Weinheim, Germany, 2009).

\bibitem{matrixbook}
R. A. Horn and C. R. Johnson, \emph{Matrix Analysis} (Cambridge University
Press, NY, USA, 1985).

\bibitem{BAmodel}
A.-L. Barab\'asi and R. Albert, Science \textbf{286}, 509 (1999).

\bibitem{ERmodel}
P. Erd\"os and A. R\'enyi, Publicationes Mathematicae \textbf{6}, 290 (1959).

\bibitem{RWLL:PRL10}
J. Ren, W.-X. Wang, B. Li, and Y.-C. Lai, Phys. Rev. Lett. \textbf{104}, 058701 (2010).

\bibitem{quantum_network}
See, for example,
M. Yanagisawa, Phys. Rev. A \textbf{73}, 022342 (2006);
S. G. Schirmer, I. C. H. Pullen, and P. J. Pemberton-Ross, Phys. Rev. A \textbf{78}, 062339 (2008);
D. Burgarth, D. D'Alessandro, L. Hogben, S. Severini, and M. Young, arXiv:1111.1475.

\end{thebibliography}

\begin{thebibliography}{01}

\bibitem{MotterComment}
J. Sun, S. P. Cornelius, W. L. Kath, A. E. Motter, arXiv:1108.5739v1 (2011).

\bibitem{graphtheory}
B. Bollob\'as, \emph{Random Graphs (2nd ed.)} (Cambridge University Press, Cambridge, UK, 2001).

\bibitem{LSB:2011nature}
Y.-Y. Liu, J.-J. Slotine and A.-L. Barab\'asi, Nature (London),
473, {\bf 167} (2011).

\bibitem{linearcontrol}
W. J. Rugh, \emph{Linear System Theory (2nd ed.)} (Prentice-Hall, NJ, USA, 1996).

\bibitem{Soptimal}
F. L. Lewis and V. L. Syrmos, \emph{Optimal Control (2nd ed.)} (Wiley, NY, USA, 1995).
\end{thebibliography}
\end{document}